\documentclass[fleqn,usenatbib]{mnras}

\usepackage{newtxtext,newtxmath}

\usepackage[T1]{fontenc}
\usepackage{ae,aecompl}

\usepackage{graphicx}	
\usepackage{amsmath}	
\usepackage{amssymb}	
\usepackage{subfig}

\usepackage{rotating}
\usepackage{adjustbox}

\title[Unwinding Spirals in GASP]{GASP XXIX - Unwinding the arms of spiral galaxies via ram-pressure stripping}

\author[C. Bellhouse et al.]{
Callum Bellhouse,$^{1,2}$\thanks{E-mail: callumb@star.sr.bham.ac.uk (CB)}
Sean L. McGee,$^{2}$
Rory Smith,$^{3}$
Bianca M. Poggianti,$^{1}$\newauthor
Yara L. Jaff\'e,$^{4}$
Katarina Kraljic,$^{5}$
Andrea Franchetto,$^{6,1}$
Jacopo Fritz,$^{7}$\newauthor
Benedetta Vulcani,$^{1}$
Stephanie Tonnesen,$^{8}$
Elke Roediger,$^{9}$
Alessia Moretti,$^{1}$\newauthor
Marco Gullieuszik,$^{1}$
Jihye Shin$^{3}$
\\
$^{1}$INAF - Astronomical Observatory of Padova, vicolo dell'Osservatorio 5, I-35122 Padova, Italy\\
$^{2}$University of Birmingham School of Physics and Astronomy, Edgbaston, Birmingham B15 2TT, England\\
$^{3}$Korea Astronomy and Space Science Institute, 776, Daedeokdae-ro, Yuseong-gu, Daejeon, 34055, Korea\\
$^{4}$Instituto de F\'isica y Astronom\'ia, Universidad de Valpara\'iso, Avda. Gran Breta\~na 1111 Valpara\'iso, Chile\\
$^{5}$Royal Observatory Edinburgh, Blackford Hill, Edinburgh EH9 3HJ\\
$^{6}$Dipartimento di Fisica \& Astronomia "Galileo Galilei", Universit\`a di Padova, vicolo dell' Osservatorio 3, 35122, Padova, Italy\\
$^{7}$Instituto de Radioastronom\'ia y Astrof\'isica, UNAM, Campus Morelia, A.P. 3-72, C.P. 58089, Mexico \\
$^{8}$Center for Computational Astrophysics, Flatiron Institute, 162 5th Ave, New York, NY 10010, USA\\
$^{9}$E.A. Milne Centre for Astrophysics, Department of Physics and Mathematics, University of Hull, Hull, HU6 7RX, UK
}

\date{Accepted 2020 October 16. Received 2020 September 4; in original form 2020 April 23}

\pubyear{2020}
\begin{document}
\label{firstpage}
\pagerange{\pageref{firstpage}--\pageref{lastpage}}
\maketitle

\begin{abstract}
We present the first study of the effect of ram-pressure "unwinding" the spiral arms of cluster galaxies. We study 11 ram-pressure stripped galaxies from GASP (GAs Stripping Phenomena in galaxies) in which, in addition to more commonly observed "jellyfish" features, dislodged material also appears to retain the original structure of the spiral arms. Gravitational influence from neighbours is ruled out and we compare the sample with a control group of undisturbed spiral galaxies and simulated stripped galaxies. We first confirm the unwinding nature, finding the spiral arm pitch angle increases radially in 10 stripped galaxies and also simulated face-on and edge-on stripped galaxies. We find only younger stars in the unwound component, while older stars in the disc remain undisturbed. We compare the morphology and kinematics with simulated ram-pressure stripping galaxies, taking into account the estimated inclination with respect to the intracluster medium and find that in edge-on stripping, unwinding can occur due to differential ram-pressure caused by the disc rotation, causing stripped material to slow and "pile-up".
In face-on cases, gas removed from the outer edges falls to higher orbits, appearing to "unwind". The pattern is fairly short-lived (<0.5Gyr) in the stripping process, occurring during first infall and eventually washed out by the ICM wind into the tail of the jellyfish galaxy.
By comparing simulations with the observed sample, we find a combination of face-on and edge-on "unwinding" effects are likely to be occurring in our galaxies as they experience stripping with different inclinations with respect to the ICM.
\end{abstract}




\section{Introduction}\label{sec:intro}

The development and evolution of galaxies is heavily influenced by gas inflows and outflows, since gas is the fuel of star formation in galaxies. Understanding how gas behaves both within the disc and in the surroundings of a galaxy is therefore vital in building a complete picture of its evolution. Galaxies can experience changes in their gas content in isolation, due to internal effects such as winds from e.g. Wolf-Rayet and OB-type stars, supernova feedback and feedback from active galactic nuclei (AGN) activity \citep{2005ARA&A..43..769V,2014MNRAS.444.3894H}. External environmental forces can also influence and, in some cases, dramatically alter the gas content within a galaxy \citep{Boselli2006}.
Within galaxy clusters, the largest gravitationally-bound objects in the universe, there are many environmental processes which can affect individual galaxies owing to the increased frequency of interactions with other galaxies, and also to interaction with the dense intracluster medium (ICM). As galaxies move through a cluster they may experience gravitational and hydrodynamical effects. Gravitational effects, such as tidal interactions with the cluster \citep{Byrd1990,Valluri1993} and gravitational interactions between galaxies \citep{Spitzer1951,Toomre1977,Tinsley1979,1983ApJ...264...24M,Mihos1993,Springel2000}, affect both the stellar and gas components of a galaxy. Hydrodynamical effects resulting from the cluster ICM affect only the gas content of the galaxy, as the stars are too massive and compact to be susceptible to hydrodynamical influence.

One particular hydrodynamical effect which plays a key role in transforming and quenching galaxies in clusters is ram-pressure stripping \citep[RPS,][]{1972ApJ...176....1G}.
In the process of RPS, the intracluster medium (ICM) interacts directly with the gas component of an infalling galaxy. The process can ultimately lead to complete removal of this gas as well as many associated observable effects including compression of gas along the leading edge of the disc \citep{Rasmussen2006,Poggianti2019b}, a boost in star formation during the peak stripping phase \citep{Vulcani2018} followed by quenching in the final stages of stripping \citep{Vulcani2020}, the formation of tails of gas in the wake of the galaxy \citep{Fumagalli2014,Cramer2019} and the condensation of star forming knots in the tails \citep{Kenney2014, Poggianti2019}.

The extended tails of jellyfish galaxies are easily seen in studies of the ionised gas, and can be narrow and collimated along the reverse of the direction of motion. \citep{Kenney1999,Fumagalli2014,Fossatti2018,Boselli2018b}

The GASP survey \citep{Poggianti2017} utilises the powerful Multi-Unit Spectroscopic Explorer \citep[MUSE:][]{2010SPIE.7735E..08B} at the Very Large Telescope in Chile to study in detail a sample of 94 ram-pressure stripped galaxies drawn from the \citet{Poggianti2016} sample. These galaxies were selected by visual inspection based on the presence of ram-pressure stripping signatures, such as displaced tails and asymmetric morphologies, from the WIde-field Nearby Galaxy-cluster Survey (WINGS) \citep{Fasano2006,Moretti2014}, its extension OmegaWINGS \citep{Gullieuszik2015,Moretti2017} and the Padova-Millennium Galaxy and Group Catalogue \citep{Calvi2011}. The aim of the GASP survey is to explore and characterise the nature of the stripping process and its effect on a multitude of galaxies in different environments. This large sample has facilitated the opportunity to observe many new aspects and effects of ram-pressure stripping, including the possibility that RPS can funnel the gas into the central region of the galaxy, triggering and fuelling the AGN \citep{Poggianti2017b} and the widespread evidence of in-situ star formation in the tails \citep{Poggianti2019}.
One of the most heavily studied jellyfish galaxies of the GASP survey to date has been JO201, analysed both individually in \citet{Bellhouse2017,George2018,George2019,Bellhouse2019,Ramatsoku2020} and as part of wider studies \citep[][]{Poggianti2017b,Vulcani2018,Moretti2018,Radovich2019}. JO201 is a massive galaxy falling into the massive cluster Abell 85 at supersonic speeds close to the line of sight toward the observer, exhibiting signs of extreme near-face-on ram pressure stripping.
The spatial distribution of the stripped knots hinted for the first time the possibility that the spiral arms  could be "unwound" or broadened as an effect of the face-on gas stripping by the ICM (see \citealt{Bellhouse2017} and Figure 2 of \citealt{George2018}). In addition to the more commonly observed "jellyfish" features such as compression of the disc and one-sided tails, the trailing knots and the ionised gas appear to be distributed along the curves of the spiral arms.

Since such observed structures in the tails resemble the tidal tails associated with gravitational interactions, it is paramount to understand the implications of such patterns being observed in galaxies undergoing ram-pressure stripping. In particular, is there a mechanism by which ram-pressure alone can lead to the "unwinding" feature, without the influence of gravitational interactions? Furthermore, since in many previous studies of jellyfish galaxies, such an effect has not been noted, is there a specific set of conditions under which this "unwinding" takes place? In comparison to tidal interactions, the unwinding effect in ram-pressure stripped galaxies appears to be accompanied by other signatures of the interaction, such as tail features, compression of the disc and the formation of trailing knots of ionised gas. Whilst unwinding alone is not an indicator of ram-pressure stripping, we aim to understand what the unwinding morphologies can tell us about a galaxy undergoing the process. Furthermore, understanding the ability of ram-pressure stripping to induce such a pattern is important when considering sample selection from broad-band data. Since curved tails are typically associated with tidal interactions, potential ram-pressure stripped galaxies may be unduly excluded by visual inspection or automated feature-detection, and in such cases further analysis may be required to constrain the true nature of the interaction. Since many studies of ram-pressure stripped galaxies, including the GASP survey, are based on visual identification of stripping features, it is crucial to ensure that this selection does not exclude a population of ram-pressure stripped galaxies which are being unwound, but show minimal to no signs of "classic" ram-pressure stripping signatures.

In the literature, a limited number of examples of "unwinding" effects in ram-pressure stripped galaxies can be found.

In \citet{Schulz2001}, hydrodynamical simulations of face-on and inclined stripping galaxies reveal the formation of flocculent spiral arms in the non-stripped disc gas as the remnant disc is compressed. Furthermore, the simulations show stretching and shearing of outer spiral arms prior to their removal from the galaxy.

In \citet{Roediger2014}, a simulated galaxy undergoing ram-pressure stripping edge-on to the ICM shows filamentary tails extending away from the disc as the spiral arm material is stripped. The galaxy was subjected to a piston-shock followed by a constant wind speed of 2000 kms$^{-1}$, resulting in the galaxy being stripped substantially, but not completely. The galaxy disc was truncated on the leading edge by the stripping and elongated on the trailing side. The authors point out that material tailing behind the galaxy is not a "tail" in the conventional sense, but remains bound to the galaxy and rotates with the original disc. The unwinding effect is visible 185Myr after initial stripping and very prominent 263Myr after initial stripping in Figure~2 of the paper.

\citet{Steinhauser2016} carried out simulations of galaxies undergoing ram-pressure stripping under different conditions, with several cluster masses and orbital trajectories. Two of the galaxies in their sample appear to show the unwinding effect (see galaxies S1 and S4, Fig~12 therein), which are both simulated to mimic galaxies undergoing extreme plunging orbits in their host clusters.

In the observational study \citet{Wolter2015}, the galaxy NGC2276 was analysed and the source of morphological disturbance was investigated. At optical wavelengths, the galaxy exhibits a similar pattern to our sample, with loose spiral arms extended to one side of the galaxy. The authors concluded from estimation of the gravitational and hydrodynamical interaction strengths and from comparison with simulations that the galaxy will be most strongly affected by ram-pressure stripping. In the study of \citet{Tomicic2018}, evidence of tidal interaction on NGC2276 was observed in the large scale gradient in the molecular gas depletion time across the disk of the galaxy. Such variations in the star formation efficiency of the molecular gas resulting from both tidal and ram-pressure interaction may also play a role in shaping the morphology of NGC2276.

In this paper, we further investigate this effect in JO201 and in the wider sample of GASP galaxies, in the quest for more evidence and a greater understanding of spiral arm unwinding during ram-pressure stripping events.

This paper is structured as follows. Section~\ref{sec:thesample} covers the preparation of the data used in this study, the selection process used to produce the sample, and the resulting galaxies used. In section~\ref{sec:orbits} we show the locations of the sample in projected position velocity phase-space, along with their tail orientations and morphologies, using these to infer the orbits and inclinations of stripping for the galaxies in the sample. Section~\ref{sec:spiralarms} shows the analysis of the MUSE data to measure the spiral arm structures and pitch angles, and to trace the distribution of different age stellar populations in the spiral arms.
In section~\ref{sec:sims} we show the analysis of simulated galaxies undergoing ram-pressure stripping with different orientations with respect to the ICM wind, comparing with observed galaxies, analysing the effect of unwinding over time throughout infall and studying the effect of stripping on the gas kinematics to understand how the unwinding patterns can arise. Finally in section~\ref{sec:discussion} we summarise the results and draw conclusions.

Throughout this paper, as in other GASP studies, we adopt a Chabrier initial mass function \citep[IMF;][]{Chabrier2003}, and a concordance $\Lambda$CDM cosmology of $\Omega_\mathrm{M}=0.3$, $\Omega_\Lambda=0.7$, $\mathrm{H}_0=70\mathrm{km}\,\mathrm{s}^{-1}\mathrm{Mpc}^{-1}$.
  
\section{The Sample and Data}\label{sec:thesample}

In order to investigate and understand the unwinding effect of ram-pressure stripping on the spiral arms of galaxies,
we drew a sample of galaxies by visual inspection from the catalog of galaxies in the GASP survey (described in section \ref{sec:intro}). The GASP observations were reduced and processed as explained in GASP I \citep{Poggianti2017}. In short, each galaxy was corrected for Milky Way dust reddening using dust values from NASA/IRSA and correcting internal dust extinction using the Balmer decrement and the \citet{1989ApJ...345..245C} extinction law. Stellar absorption was taken into account using the spectral fitting code SINOPSIS \citep{Fritz2017} to estimate the contribution from the stellar component, which was then subtracted from each spaxel.
The data were then analysed using \textsc{kubeviz} \citep{Fumagalli2014,Fossatti2018} as detailed in \citet{Bellhouse2017}. \textsc{kubeviz} fits a series of selected emission lines on a spaxel-by-spaxel basis with 1 or 2 gaussian component profiles, allowing resolved analysis of the distribution and kinematics of the major emission lines in the observation to be carried out. We used \textsc{kubeviz} on datacubes after smoothing with a $5\times5$ kernel, with the redshift set to the mean redshift for each galaxy and noisecube errors scaled to give a reduced $\chi^2=1$. The resulting fits to the emission lines across each galaxy gave the flux and velocity measurements used in this analysis. The stellar velocities for each galaxy were calculated using the absorption lines with the \textit{ppxf} code \citep{Cappellari2004} as described in \citet{Poggianti2017}. 
%
%
The sample of unwinding galaxies, i.e. galaxies which exhibit visual evidence of unwinding spiral arms, was selected from the full GASP catalogue of 94 jellyfish galaxies by visual inspection of the whitelight MUSE images along with the $\mathrm{H}\alpha$ maps created using \textsc{kubeviz}.

The visual selection was carried out with the aim of singling out objects with the following characteristics:
\begin{enumerate}
    \item Disc morphology (excluding mergers, rings, irregular galaxies etc).
    \item Orientation close to face-on, (axial ratio $> 0.5$). This criteria ensures that the state of the spiral arms can be clearly inspected, which is not the case in edge on galaxies. 
    \item  Clear signatures of loose spiral arms, as judged from visual inspection of the  H$\alpha$ maps and whitelight images.
\end{enumerate} 



\begin{figure*}
    \centering
    \includegraphics[width=\textwidth]{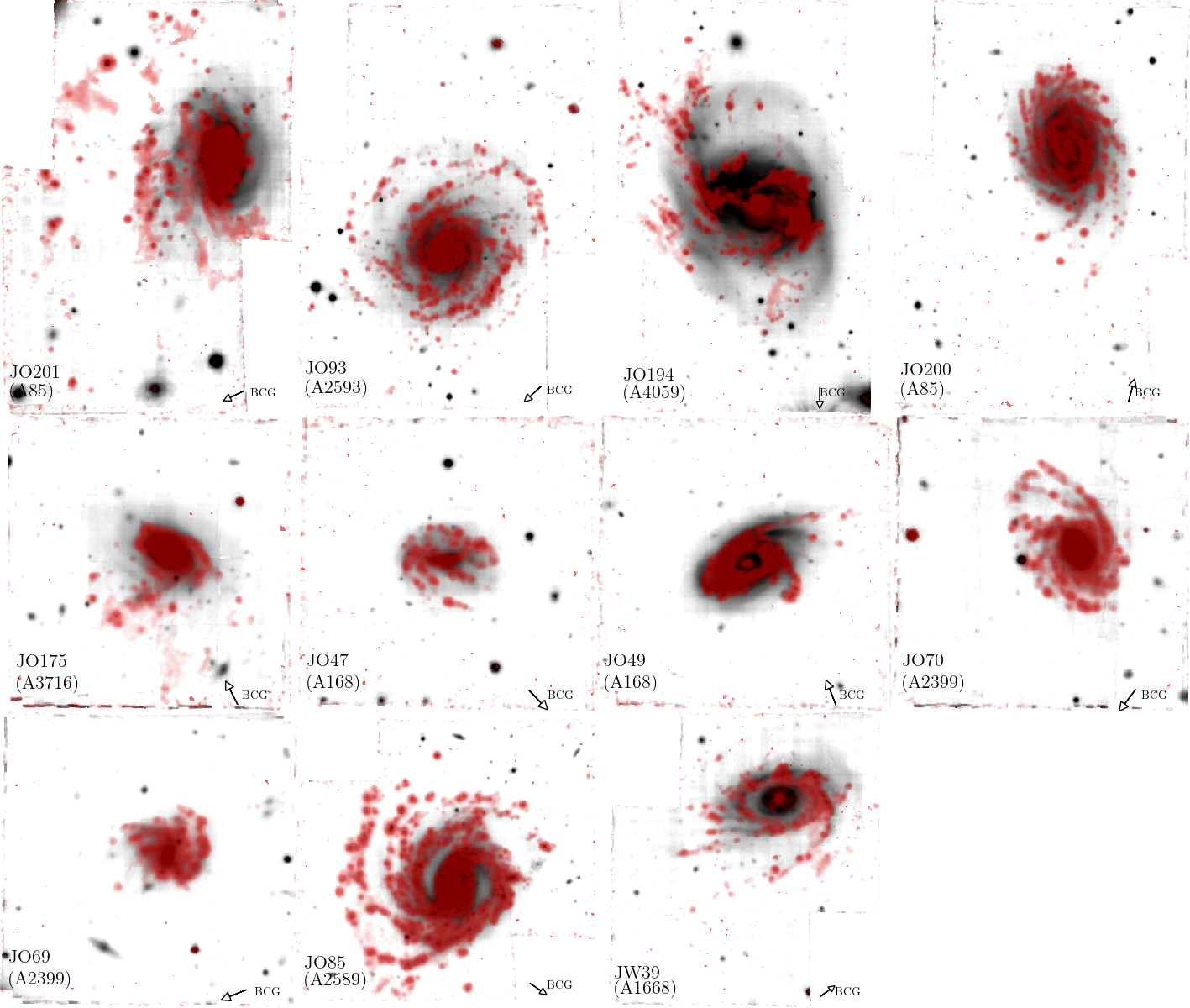}
    \caption[The unwinding galaxy sample, showing inverted whitelight images of each galaxy with the H$\alpha$ emission line flux overlaid. The figure shows the H$\alpha$ tails extending far beyond the discs in some cases, whilst maintaining a clear spiral arm pattern.]{The unwinding galaxy sample, showing inverted whitelight images of each galaxy with the H$\alpha$ emission line flux overlaid in red. The figure shows the red H$\alpha$ tails extending far beyond the discs in some cases, whilst maintaining a clear spiral arm pattern. The white light images generally show little to no unwinding pattern, which is expected to only manifest itself in the stripped gas and young stellar population.}
    \label{fig:thumbnails}
\end{figure*}

\begin{table*}
\centering
\caption[Sample of GASP jellyfish galaxies with unwinding spiral arms.]{Sample of GASP jellyfish galaxies with unwinding spiral arms.}
\begin{tabular}{l l c c c c c c}\label{table:unwinding_sample}
\textit{Stripped Galaxies}
\\\hline
Galaxy &Cluster &R.A. &Decl. & $M_{*,\rm disc}$& Inclination & Angle to & Close to face-on  \\ &&(J2000)&(J2000)&($10^{10}M_\odot$)&&LOS motion & Stripping \\
\hline
JO201	&A85	&00:41:30.29	&-09:15:45.900 &$6.2 \pm 0.8$& $41.4^\circ$ & $13^\circ \substack{+21 \\ -6}$ &\checkmark \\
JO93	&A2593	&23:23:11.74	&+14:54:05.013 & $3.5\pm0.5$ & $25.0^\circ$ & $7^\circ \substack{+7 \\ -3}$ &\checkmark \\
JO194	&A4059	&23:57:00.68	&-34:40:50.117 & 15.0$\pm$3 & $38.3^\circ$ & $8^\circ \substack{+16 \\ -5}$ &\checkmark \\
JO200	&A85	&00:42:05.03	&-09:32:03.841 & $7\pm1$ & $45.3^\circ$ & $34^\circ \substack{+15 \\ -14}$ & \\
JO175	&A3716	&20:51:17.60	&-52:49:21.825 & $3.2\pm0.5$ & $43.2^\circ$ & $61^\circ \substack{+9 \\ -15}$ & \\
JO47	&A168	&01:15:57.67	&+00:41:35.938 & $0.40\pm0.06$ & $44.7^\circ$ & $21^\circ \substack{+22 \\ -10}$ &\checkmark \\
JO49	&A168	&01:14:43.85	&+00:17:10.091 & $4.8\pm0.6$ & $53.6^\circ$ & $86^\circ \substack{+2 \\ -3}$ & \\
JO70	&A2399	&21:56:04.07	&-07:19:38.020 & $2.9\pm0.6$ & $39.3^\circ$ & $85^\circ \substack{+3 \\ -5}$ & \\
JO69	&A2399	&21:57:19.20	&-07:46:43.794 & $0.8\pm0.2$ & $43.3^\circ$ & $26^\circ \substack{+16 \\ -11}$ &\checkmark\\
JO85	&A2589	&23:24:31.36	&+16:52:05.340 & $4.6\pm0.9$ & $24.9^\circ$ & $30^\circ \substack{+15 \\ -11}$ &\checkmark\\
JW39	&A1668	&13:04:07.71	&+19:12:38.486 & $17\pm3$ & $52.5^\circ$ & $31^\circ \substack{+15 \\ -15}$ & \\ 
\hline

\\\textit{Control Galaxies}
\\\hline
Galaxy &        &R.A. &Decl. & $M_{*,\rm disc}$& Inclination &&  \\ &&(J2000)&(J2000)&($10^{10}M_\odot$)&&& \\
\hline
P21734	&	&11:31:07.90    &-00:08:07.914  &$ 6\pm1 $& 32.1$^\circ$ & \\
P25500	&	&11:51:36.28	&+00:00:01.929  &$ 7\pm1 $& 50.0$^\circ$ & \\
P48157	&	&13:36:01.59	&+00:15:44.696  &$ 3.9\pm0.8 $& 49.2$^\circ$ & \\
P19482	&	&11:22:31.25	&-00:01:01.601  &$ 2.2\pm0.4 $& 54.9$^\circ$ & \\
P63661	&	&14:32:21.79	&+00:10:41.628  &$ 2.1\pm0.3 $& 42.4$^\circ$ & \\
P95080	&	&13:12:08.70	&-00:14:20.510  &$ 1.3\pm0.5 $& 43.1$^\circ$ & \\
\hline
\end{tabular}
\end{table*}



Out of 12 face-on observed galaxies undergoing RPS with clear spiral arms in the full GASP sample, 11 met our criteria and exhibited signs of possible unwinding.
The high incidence indicates that the effect could be prevalent in stripped galaxies. The one galaxy inclined face-on to the observer not showing the unwinding effect is galaxy JO128 which is located towards its cluster outskirts and belongs to the control sample of non-stripped galaxies given in \citet{Vulcani2018} due to the low star formation measured in the galaxy's tails.

In order to rule out gravitational interactions within the sample, a companion search was carried out using the photometric and spectroscopic data from OmegaWINGS. Criteria for the search were:
\begin{enumerate}
    \item Photometric companions with V-band magnitude < 20, at least 3 magnitudes deeper than the faintest galaxy in the sample and within $20\times\mathrm{R}_\mathrm{e}$ of the galaxy.
    \item Spectroscopic companions with distance $<20\times\mathrm{R}_\mathrm{e}$
\end{enumerate}
Using these criteria, 7/11 galaxies have no discernible companions which could be a source of gravitational influence. The 4 galaxies with possible neighbours are: 1) JO201, whose neighbours are $>10\times\mathrm{R}_\mathrm{e}$ and 2 of which have large peculiar velocities. 2) JO194, with one neighbour at least 5 Magnitudes fainter than itself. 3) JO69, with a more compelling companion ~57kpc from the galaxy, with similar velocity and magnitude. 4) JO200, with three spectroscopic companions, all ~3 magnitudes fainter and outside $10\times\mathrm{R}_\mathrm{e}$.

Out of these, JO69 is the most likely to be affected by gravitational influence, whilst the companions of the other three galaxies are unlikely to be massive enough or close enough to cause any significant disturbance.

The final sample of unwinding spiral galaxies is listed in Table~\ref{table:unwinding_sample} and shown in Figure~\ref{fig:thumbnails}.
Additionally to the unwinding sample, we selected a control sample of galaxies from the GASP catalog \citep[see][for the full GASP control sample]{Vulcani2019}. The selected control galaxies are not undergoing stripping and have face-on orientations, disc morphologies and clear spiral arms as with the unwinding sample. The six control galaxies are also presented in Table~\ref{table:unwinding_sample}. We note that two of the control sample galaxies, P63661 and P95080, are suspected to be experiencing cosmic web enhancement \citep{Vulcani2019}, but our results do not suggest that this is affecting the morphology of these galaxies.

\subsection{Morphologies}

As Figure~\ref{fig:thumbnails} shows, the effect of unwinding spiral arms is most visible in the $\mathrm{H}\alpha$ emission (not in the stellar light).
Notable examples of unwinding spiral arms include JO201, which in the $\mathrm{H}\alpha$ map shows an extremely extended arc of stripped material to the east of the galaxy in Figure~\ref{fig:thumbnails}, far outside the galaxy disk. JO200 shows a generally uniform spiral morphology, with three spiral arms extending below the disk in the image. While some galaxies exhibit asymmetric unwinding patterns in a distinct tail, JO85 and JO93 show an unwinding pattern most of the way around the disc. It is possible that the azimuthal extent of the unwinding feature around the galaxy disc may be proportional to the inclination of the galaxy with respect to the ICM, i.e. tighter tails are observed in galaxies moving edge-on into the wind, which unwind only on their trailing edges, whereas broader unwinding tails may indicate face-on stripped galaxies, whose tails originate from further around the disc.

All of the galaxies in the sample have generally undisturbed inner disks which further diminishes the likelihood of any gravitational influence. In addition to the unwinding arms, all of the galaxies in the sample exhibit at least some other signatures of ram-pressure stripping, including extended extraplanar material, trailing clumps of star formation and compression of the leading edge of the disc.

\subsection{Kinematics}
\begin{figure*}
    H$\alpha$\hspace{240pt}Stars\hspace{10pt}\hphantom{1}
    \centering
    \includegraphics[width=\textwidth]{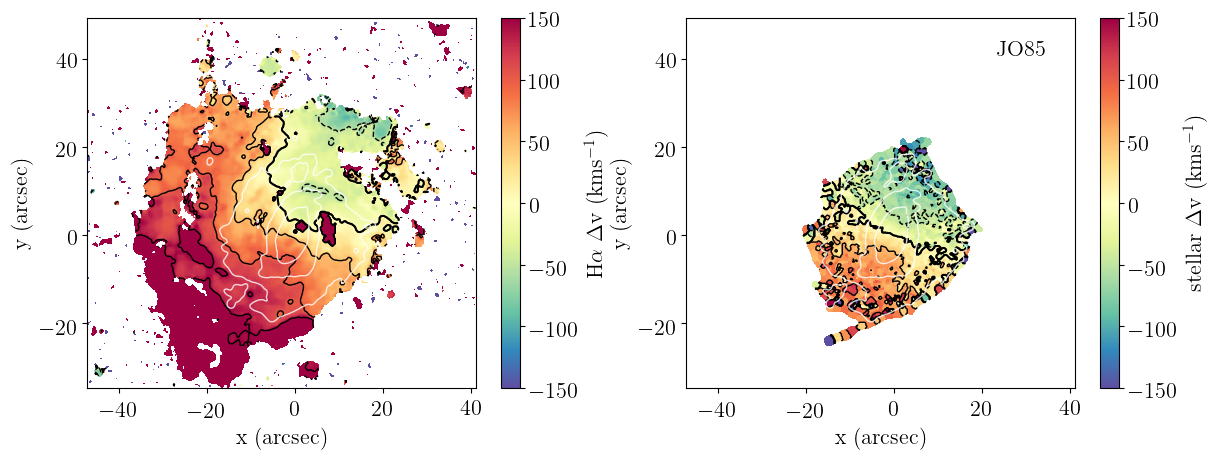}
    \includegraphics[width=1\textwidth]{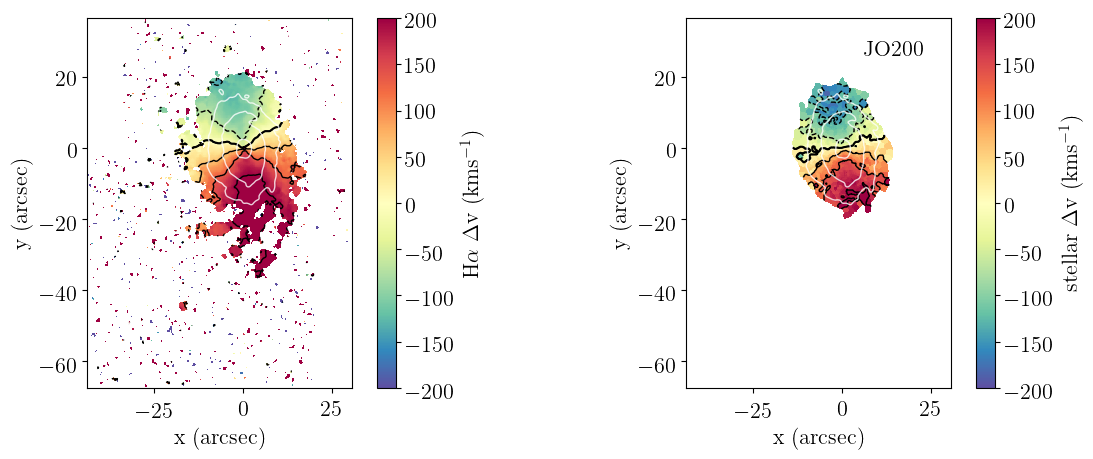}
    \caption[Comparison between ionised gas and stellar kinematics maps in JO85 and JO200.]{Comparison of kinematics maps in JO85 and JO200. Black lines indicate iso-velocity contours on the corresponding maps. The left panels show the ionised gas velocity maps for each galaxy measured from emission line kinematics. The right panels show the stellar velocity maps measured using ppxf. \textit{Top,} JO85: The stellar velocity contours exhibit a "U" shaped pattern at the upper and lower edges of the disc, in contrast, the gas kinematics follow concentric contours outward from the upper edge of the disc. The gas kinematics show a slight systematic offset to the stellar kinematics as seen by the shift of the zero contour from the centre of the galaxy disc.
    \textit{Bottom, }JO200: The stellar and gas velocities are similar in general, with some differences in the central gradient.}
    \label{fig:kin_compare_vels}
\end{figure*}

Figure \ref{fig:kin_compare_vels} compares the ionised gas kinematics on the left panels with the stellar velocities on the right, for 2 example galaxies (JO85 and JO200). These galaxies were chosen in particular as archetypal face-on and edge-on stripped galaxies respectively, to compare the effect of both scenarios. The remainder of the sample of unwinding galaxies is shown in the appendix (figure~\ref{fig:vel_compare_all}).

The stellar velocities in both galaxies are much smoother and show mostly undisturbed rotation. In the ionised gas of JO85, the kinematics are similar to the central region of the stellar disc, and significantly different around the edges. Whilst the stellar disc velocity contours are smoother and straighter at the northern and southern edges of the disc, the gas kinematics contours bend up to the right at the northern edge of the disc. This is likely to result from JO85's mostly face-on interaction with the ICM (see section \ref{sec:orbits}), pushing the gas around the edges of the disc to higher line-of-sight velocities. In the case of JO200, the gas velocities are similar to the stellar velocities, with some differences in the central gradient. The stripped ionised gas appears to continue rotating with similar velocity to the galaxy disk.

The smooth stellar kinematics in the two examples and the rest of the sample shown in figure~\ref{fig:vel_compare_all} of the appendix, combined with the stellar morphology vs $\mathrm{H}\alpha$, confirm that the disturbances experienced by these galaxies are hydrodynamical in nature, as gravitational interactions would influence both the gas and stellar kinematics.

\section{Galaxy orbits}\label{sec:orbits}

In order to explore the orbits of the galaxies in the sample and the inclination angles between the galaxies and their directions of motion through the ICM, we looked at the morphologies of galaxies and their positions in projected phase space.

\begin{figure*}
\includegraphics[width=\textwidth]
    {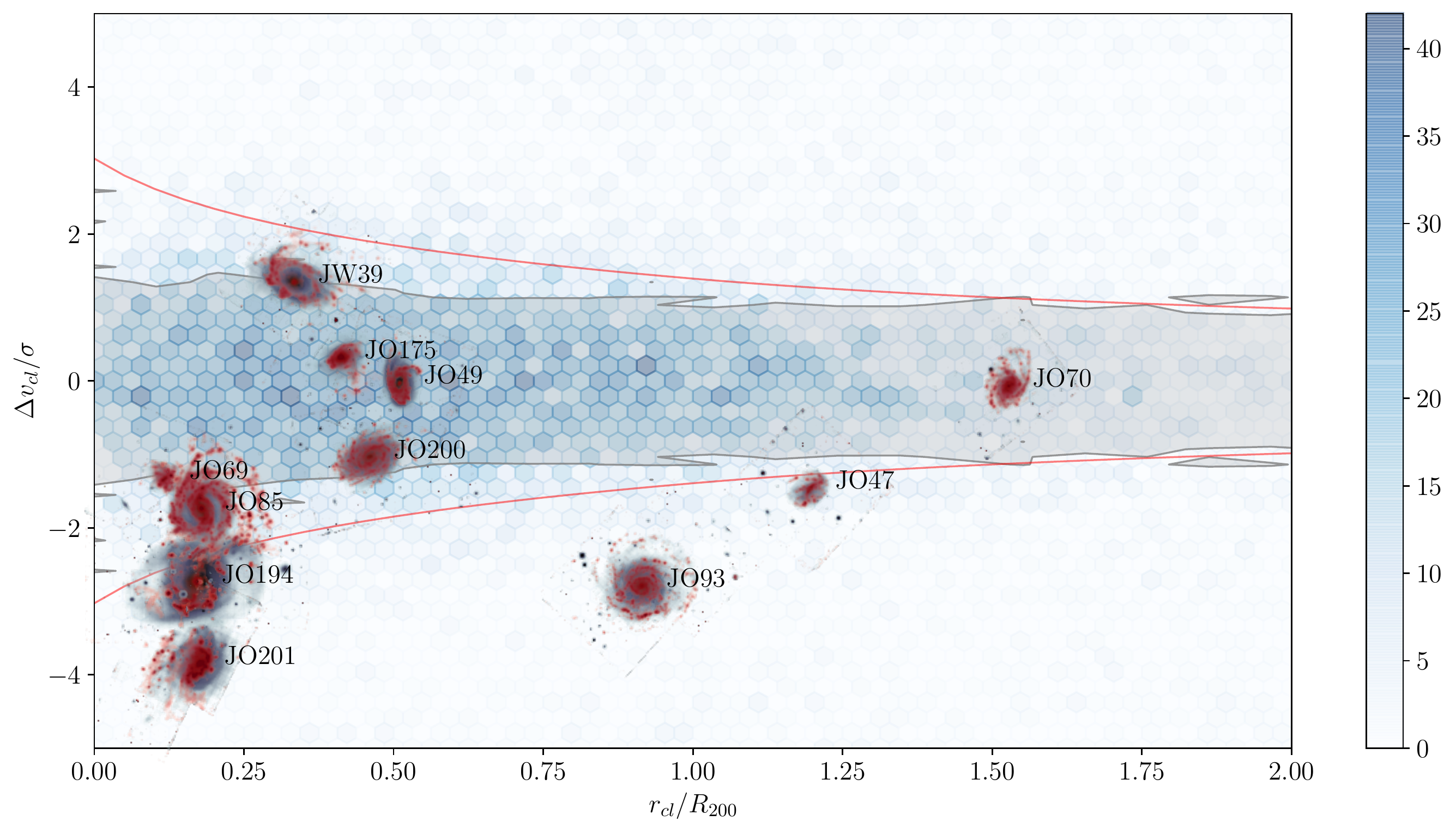}
    \caption{Projected position vs. velocity phase-space diagram of the unwinding sample, showing thumbnail images of each galaxy (Colours as in Figure~\ref{fig:thumbnails}). Each thumbnail is rotated so that the BCG direction is to the left. The sample of all omegaWINGS \citep{Gullieuszik2015,Moretti2017} galaxies is shown in the background as the blue hex plot and the solid red lines indicate escape velocity curves in a NFW profile as in \citet{Jaffe2018}. The grey shaded area encloses the region in which the mean velocity angle is greater than $30^\circ$ to the line-of-sight. Galaxies outside this region are most likely to be moving mostly toward or away from the observer.}\label{fig:phase_thumbs}
    \centering
\end{figure*}

The location of the sample of unwinding galaxies across projected position vs. velocity phase-space is a useful parameter in this analysis, as this diagram can reveal the orbital histories of the cluster galaxies \citep[see][]{Jaffe2018}. Figure~\ref{fig:phase_thumbs} shows the phase-space distribution of the unwinding-spiral jellyfish sample compared with the overall population of cluster galaxies from WINGS/OmegaWINGS. The galaxies have been plotted as thumbnail images centred on their corresponding locations in phase space. For each galaxy, the $\mathrm{H}\alpha$ map is plotted in red over the whitelight image. The thumbnail images have been rotated such that the BCG direction is always to the left, parallel to the x-axis. 

Almost all the galaxies in our sample have negative peculiar velocities which indicate they are moving towards the observer. In particular, there are several galaxies (most notably JO201, JO194 and JO85) occupying the lower envelope of the trumpet-shape distribution of galaxies in phase-space close to the cluster centre, which is an indication that they could be crossing the dense core of the ICM at high speeds on radial orbits along our line of sight on first infall \citep[see][]{Bellhouse2017,Jaffe2018}. Only one galaxy has a comparatively high positive velocity at a similar clustercentric distance (JW39), which indicates it is moving away from the observer.



According to the phase space diagram, the unwinding sample does not occupy a single, specific region of phase space, but none are located in the virialised region of phase space (low velocity, low projected radii).


The grey shaded region and grey lines indicate the probable angle of a galaxy's velocity between line-of-sight and plane-of-the-sky, (where $0^\circ$ corresponds to motion exclusively along the line-of-sight and $90^\circ$ to exclusively on the plane of the sky.) given its location in projected phase space. These were produced by analysing a sample of 42 simulated galaxy clusters (M$_{200}>1\times 10^14$ M$_\odot$). The N-body simulations used were run using GADGET \citep{Springel2005} with a box-length of the volume of 120 Mpc/h box. Initial conditions were built for a redshift=200 using the publicly available package MUSIC \citep{Hahn2011}, applying a camb power spectrum \citep{Lewis2000}, accurate to second order. The mass of a dark matter particle is 1.072e9 M$_\odot$/h. Halo finding was conducted using ROCKSTAR \citep{Behroozi2013} for all halos down to $1\times10^{11}$M$_\odot$/h. Each simulated cluster was observed from 50 random lines of sight and stacked to produce a phase space diagram of normalised projected radius and relative velocity. For each simulated cluster galaxy, the angle between its velocity vector and the line of sight axis was measured, and across the grid of phase space, these values were binned to give the median velocity angle for any given location. The galaxies in our observed sample were then assigned the velocity angle corresponding to their measured location in phase space. Further details of the simulations and method will be presented in an upcoming dedicated paper (Smith et al. in prep.).

Obtaining an estimate of the velocity angle for each galaxy allows us to more reliably predict whether a given galaxy is moving mostly along the line-of-sight, or across the plane of the sky. For a galaxy viewed close to face-on, this yields the inclination of the galaxy with respect to the ICM wind, and therefore a measure of whether the galaxy is experiencing face-on or edge-on RPS. Since many galaxies in our sample are not viewed fully face-on, it is hard to predict the true wind-angle due to uncertainties in the alignment of the galaxy's observed inclination and the true direction of motion along the plane of the sky. We can, however, use this as an indication of the likelihood of face-on stripping, which is more easily constrained.
Galaxies viewed face-on and located outside the grey shaded area are therefore most likely to be experiencing face-on stripping. The values of the predicted angle to LOS motion are given in Table~\ref{table:unwinding_sample}, as well as the likelihood that the velocity is within $30^\circ$ of the line-of-sight.

Out of the sample of 11 unwinding galaxies, 6 galaxies, JO201, JO93, JO194, JO85, JO69 and JO47 are all outside the grey region in phase space, and are moving within $30^\circ$ of the LOS. Coupled with their low inclinations, this suggests that these galaxies may be moving mostly face-on with respect to the ICM. The remaining 5 galaxies are likely to be undergoing either mostly edge-on stripping or a combination of the two.



The thumbnail images in the diagram give an indication of the projected tail directions of each of the unwinding galaxies with respect to the cluster centre. The general orientation of the tail toward the left or right (toward or away from the BCG, respectively) can indicate a galaxy's progression through infall, as the tail will be pointed away from the cluster centre up until pericentric passage and toward the cluster centre after pericentre, however for the galaxies moving along the line of sight, this is less well defined. The galaxies with more asymmetric tails, which lie to one side of the disc, are generally found within the grey shaded region, where they are likely to be experiencing some degree of edge-on stripping.
Meanwhile the galaxies with shorter, broader tails, which in some cases appear to be more symmetrically unwinding, are generally located within the "face-on" stripping region at high LOS velocities, outside the grey region.

\section{Spiral Arms}\label{sec:spiralarms}


\subsection{Identifying and characterising spiral structure}\label{sec:azimuthal}


In order to analyse the extent of the unwinding spiral arm patterns in the galaxy sample, we produced maps of the $\mathrm{H}\alpha$ emitting ionised gas,
reprojected onto polar space in terms of radial distance from the centre of the galaxy vs azimuthal angle around the disc.

The tightness of a spiral arm of a galaxy is measured by the "pitch angle", which refers to the angle between a spiral arm and the tangent to the circle on the plane of the disc, where zero lies along the tangent and 90$^\circ$ lies along the normal. More tightly wound spiral arms have lower pitch angles whilst steeper, more loosely wound spiral arms have higher pitch angles. We thus measured the pitch angles to quantify the unwinding by comparing inner and outer regions of the galaxy disks, in both the stripped and control sample galaxies. Higher pitch angles in the outer regions of the disk compared with the inner disk would reveal that the spiral arms are effectively being unwound.

We chose to implement a manual method of measuring the pitch angles of the spiral arms in this analysis. In particular this is because these galaxies, by selection, show very disturbed morphologies which are not easily fitted using an automated process. In some cases, particularly JO201, the stripped material is so far removed from the galaxy that it barely qualifies as a conventional "spiral arm", but its shape and alignment certainly indicate that it is likely to have originated from one. In other cases, the spiral arms have been greatly extended by stripping to one side of their galaxy, and compressed or removed on the other. In order to make use of these extreme cases, a manual method allowed more fine-tunable selection of the material used in defining each spiral arm, ensuring more accurate reproduction of the local pitch angles.
Since the sample of unwinding galaxies shown here is fairly small, a manual method of fitting the spiral arms is also within the realm of practicality.

\begin{figure*}
    \centering
    \includegraphics[width=\textwidth]{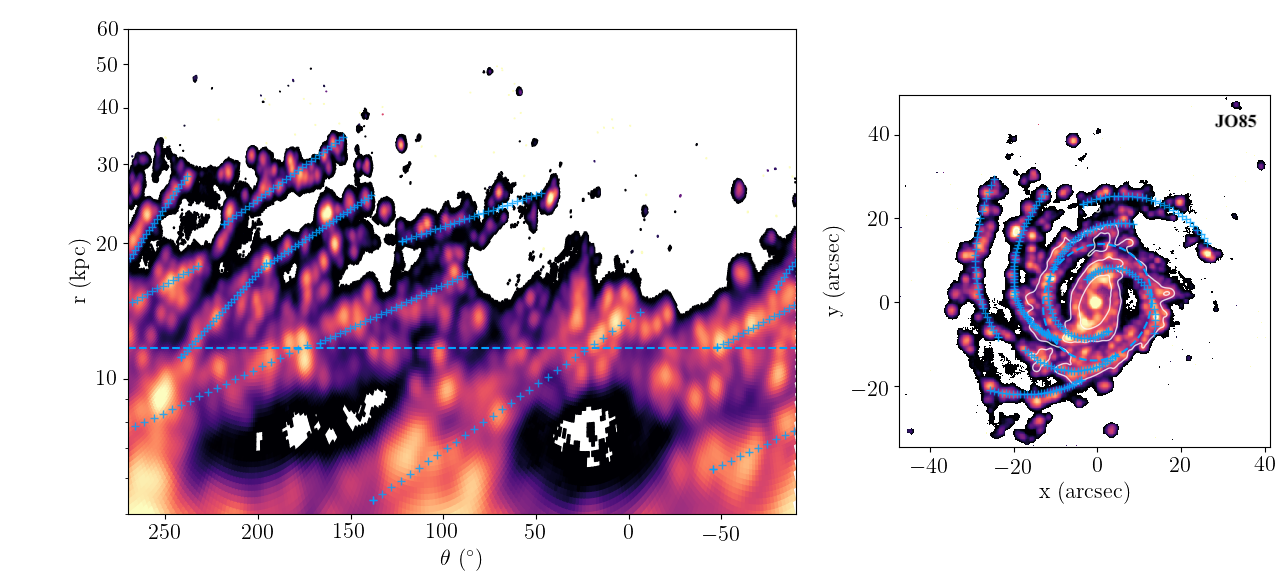}
    \includegraphics[width=\textwidth]{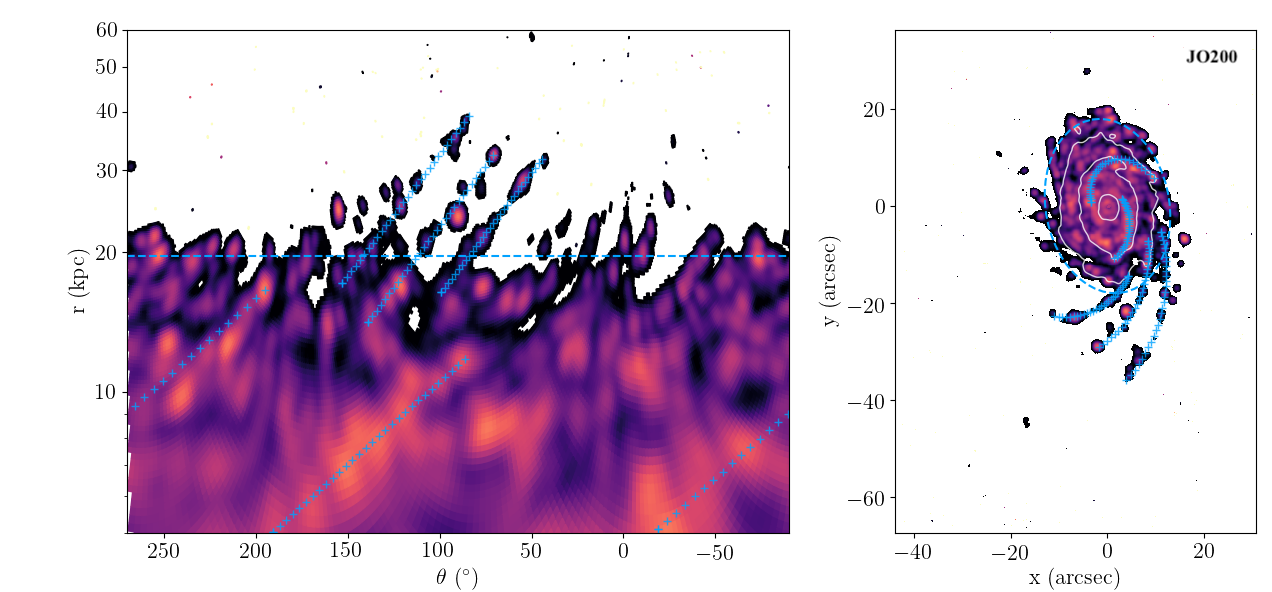}
    \caption[Plots of H$\alpha$ for the "unwrapped" disc of JO85 and JO200, showing radial projected distance from the centre against azimuthal angle.]{Azimuthal plots of H$\alpha$ for the "unwrapped" disc of JO85 (\textit{Top}) and JO200 (\textit{Bottom}), showing radial projected distance from the centre in logarithmic scale against azimuthal angle in the left panels, alongside the original galaxy discs on the right panels. The white contours on the right panels denote stellar continuum isophotes. The blue dashed horizontal lines on the left panels and the dashed blue ellipses on the right mark 2 effective radii on the disc. Lines of + symbols mark spiral arm patterns identified by eye on the unwrapped figure and are shown reprojected back on the original galaxy discs. The pitch angles of these spiral arms are shown along with their average radial distance in Figure~\ref{fig:pitchangles}.}
    \label{fig:JO85_unwrap}
\end{figure*}

\begin{figure*}
    \centering
    \includegraphics[width=\textwidth]{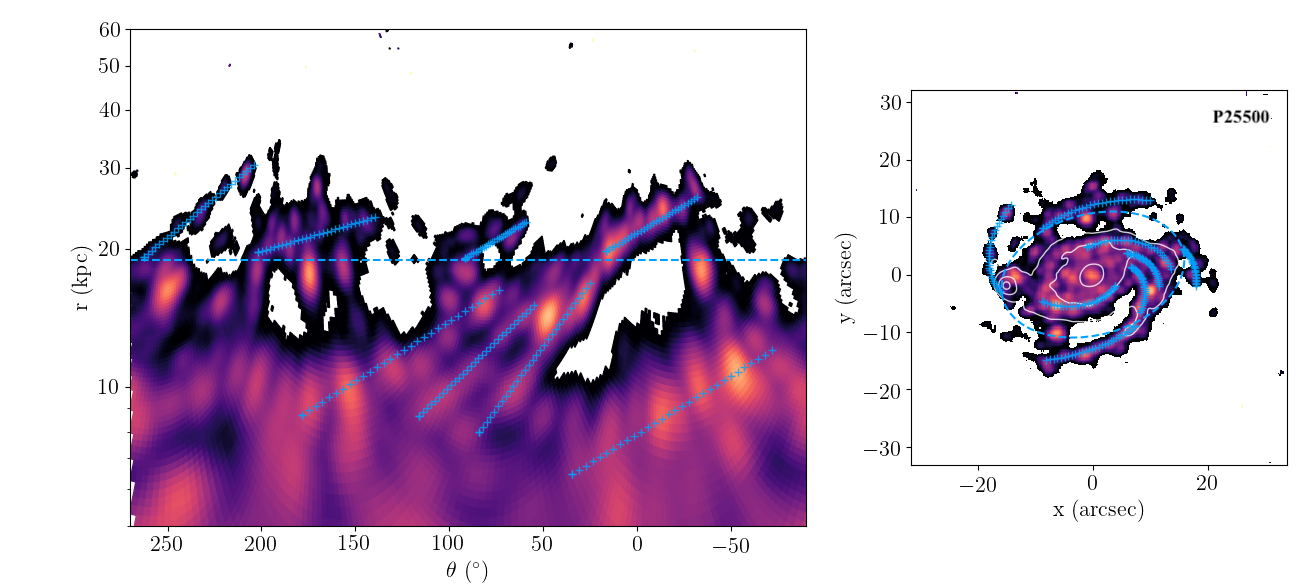}
    \caption[Plot of H$\alpha$ for the "unwrapped" disc of P25500, showing radial projected distance from the centre against azimuthal angle.]{Plot of H$\alpha$ for the "unwrapped" disc of P25500, showing radial projected distance from the centre in logarithmic scale against azimuthal angle, alongside the original galaxy disc on the right. The white contours on the right panel denote stellar continuum isophotes. The dashed blue line in on the left panel and the dashed blue ellipse on the right both indicate 2 effective radii on the disc. Lines of "+" symbols mark spiral arm patterns identified by eye on the unwrapped figure and are shown reprojected back on the original galaxy disc. The pitch angles of these spiral arms are shown along with their average radial distance in Figure~\ref{fig:pitchangles}.}
    \label{fig:P25500_unwrap}
\end{figure*}

We use the $\mathrm{H}\alpha$ emission, as it traces the distribution of the star formation activity which is increased in the denser regions of gas and dust of the spiral pattern.
The emission maps were taken directly from the \textsc{kubeviz} fits.
In order to carry out the reprojection, each galaxy was corrected for inclination according to the axial ratios measured in \citet{Franchetto2020}, by scaling the distances along the dimension of the kinematic minor axis. Each spaxel was then reprojected onto polar coordinate space according to its radial distance and azimuthal position. The reprojection was carried out by calculating the position of each corner of each spaxel and transforming into the reprojected polar space, in order to preserve the extent of each spaxel in the new space, particularly in regions close to the galaxy centre where individual spaxels cover a large range of azimuthal angles.
Figure~\ref{fig:JO85_unwrap} shows example figures for JO85 and JO200 (same galaxies shown in figure~\ref{fig:kin_compare}) where the disc of the galaxy is displayed "unwrapped" in the left panel, alongside the original galaxy on the right. Logarithmic spiral arms with constant pitch angle in the galaxy appear as lines on the left hand side, with the gradient of the lines indicating the pitch angle. A galaxy from the non-stripping control sample, P25500, is shown in Figure~\ref{fig:P25500_unwrap} for comparison.

Straight lines were aligned with the spiral arms on the azimuthal figure by eye, and the reverse of the reprojection method was used to show these on the original galaxy, highlighting the arm locations.

In figure~\ref{fig:JO85_unwrap}, the unwinding effect manifests itself as a steepening of the pitch angle in the stripped material. In most of the strongly unwinding galaxies, the steepening occurs not gradually but sharply, with distinctly differing pitch angles between the disk and stripped material, however this steepening occurs at a different point for each spiral arm.

The pitch angle of each spiral arm was taken as the mean value of 4 measurements, independently repeated. This was then averaged between all spiral arms to obtain the global mean pitch angle within the disc (defined here as having average radii within $2 \times$ the effective radius and within the stellar continuum contours) and in the outer disc, for each galaxy.

The effective radii were calculated for each galaxy by analysing the azimuthally averaged surface brightness profile in I-band images generated from the MUSE data \citep{Franchetto2020}.

The pitch angles were also measured for the stellar continuum emission using the same method (not shown here). For the unwinding sample, these were all in agreement with the inner disc H$\alpha$ pitch angles to within $10^\circ$, with an average offset of $4.8^\circ$, indicating that the inner spiral arm pitch angle is a reasonable measure of the "undisturbed" pitch angle.

In the case of JO85, the spiral arms have a mean pitch angle of $17.3^\circ$ in the disc, which increases to around $23.0^\circ$ in the unwound tails. For JO200, the spiral arms inside and outside 2 effective radii have mean values of $22.0^\circ$ and $30.5^\circ$ respectively. In contrast, the control galaxy shown here, P25500, has mean pitch angles of $24.8^\circ$ and $18.6^\circ$ inside and outside 2 effective radii, respectively.

The azimuthal plots of other galaxies, shown in Appendix \ref{azimuthal_sample}, Figure~ \ref{fig:azimuthal_compare_all}, have average disc and tail spiral arm pitch angles shown in Table~\ref{table:pitchangles}. Cases where spiral arms are not clear enough to measure an accurate pitch angle are indicated by question marks. Azimuthal projections of the simulated galaxies (see section~\ref{sec:sims}) were also produced using the same technique for comparison, and are shown in figure~\ref{fig:sim_unwrap} of the Appendix.

\begin{figure*}
    \centering
    Unwinding Sample\hspace{200pt}Control Sample
    \includegraphics[width=0.49\textwidth]{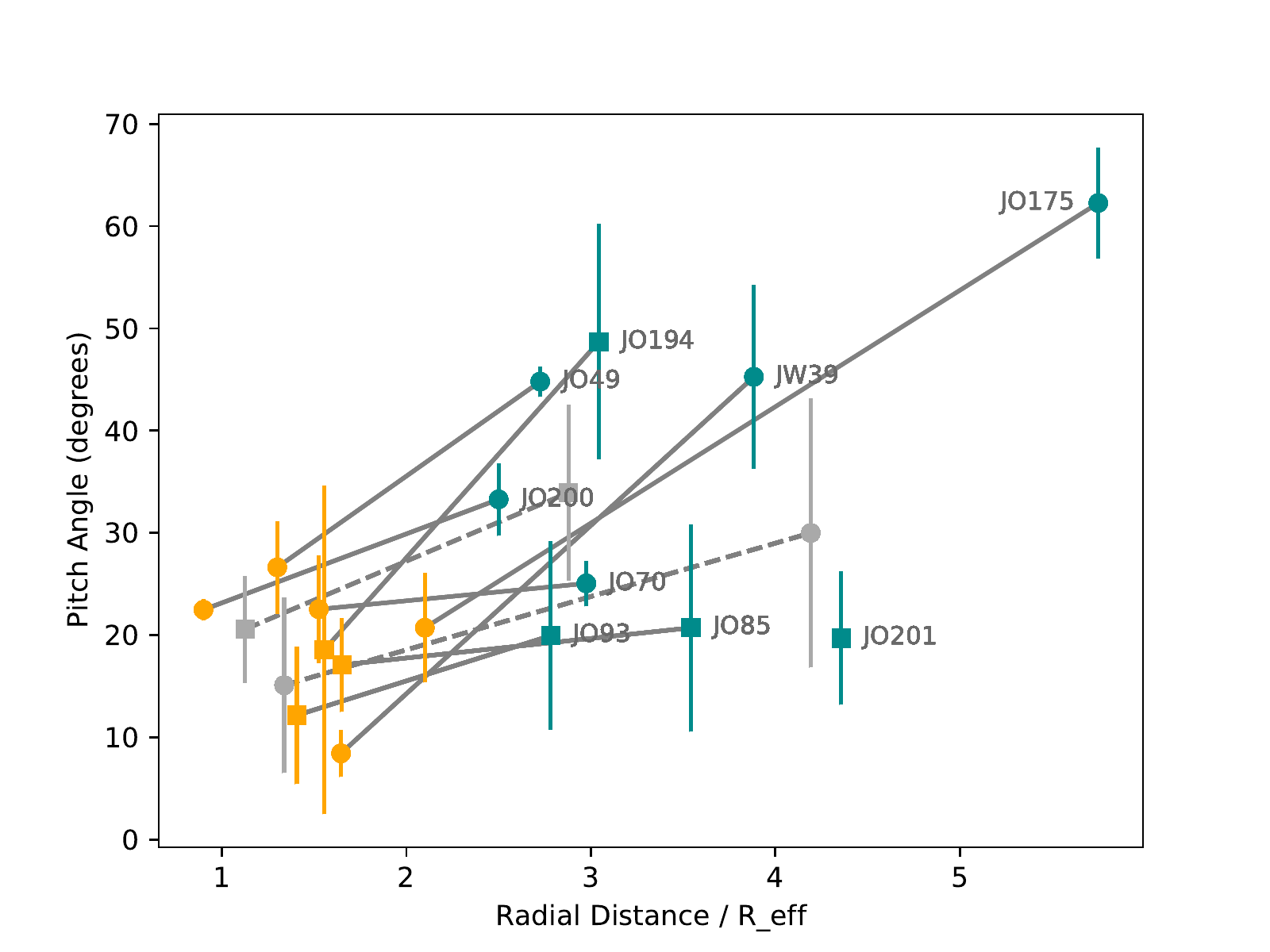}
    \includegraphics[width=0.49\textwidth]{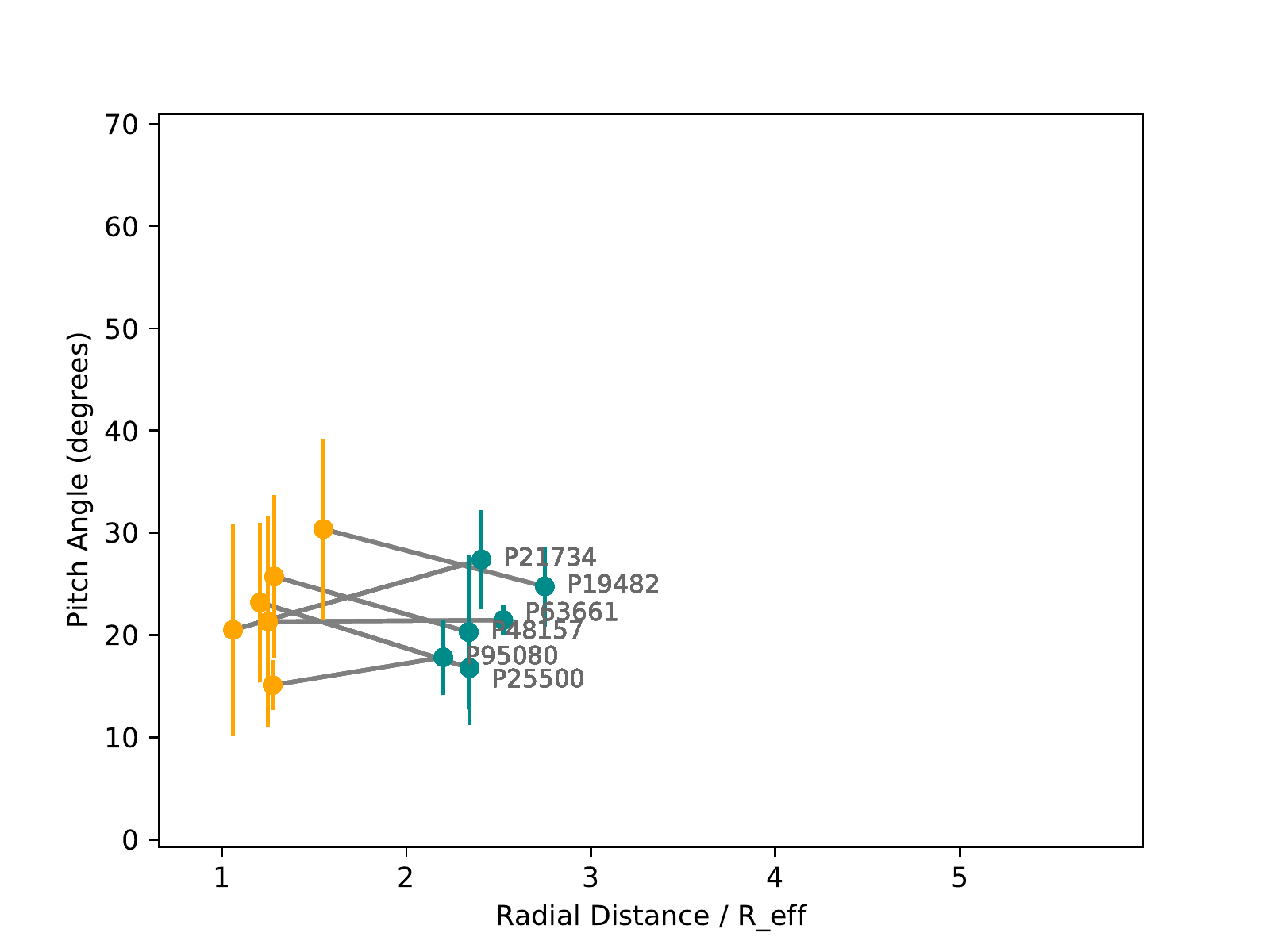}
    \caption[]{Measured average pitch angles for all galaxies in the unwinding sample (\textit{Left}) and control sample (\textit{Right}). "Inner" spiral arms, marked in orange, are defined where the average radial distance is within 2 r$_{\rm eff}$ and the arm lies within the visible stellar continuum contours (examples shown in white in the right panels of figure~\ref{fig:JO85_unwrap}). Outer spiral arms are marked by teal points. Galaxies tagged as face-on interacting in Table~\ref{table:unwinding_sample} are shown as square markers. The values are connected by grey lines to indicate which points belong to the same galaxy, and errorbars mark the standard deviation between the measured pitch angles. The grey points connected by dashed lines indicate the simulated face-on and edge-on stripped galaxies, the azimuthal plots for which are shown in figure~\ref{fig:sim_unwrap} of the Appendix.}
    \label{fig:pitchangles}
\end{figure*}

To measure and characterise the extent of "unwinding" in the outer spiral arms, the pitch angles for all galaxies in the unwinding and control samples, as well as the simulated unwinding galaxies, are shown in Figure~\ref{fig:pitchangles} against their mean radial distance in effective radii. For each galaxy, the mean inner and outer spiral arm pitch angle and radius values are connected by grey lines for comparison. The left hand panel of Figure~\ref{fig:pitchangles} shows the unwinding sample, while the right hand panel shows the control galaxies. The outer spiral arms of the control galaxies extend much less far than the unwinding galaxies. The ranges of masses and redshifts of the control galaxies are the same as those of the stripped sample, therefore the lack of spiral arms at higher radii in the control galaxies is due to the absence of stripped material only. It is clear from the figure that the control galaxies have much lower outer spiral arm pitch angles compared to the stripped galaxies, and the gradients between inner and outer pitch angles are generally low, indicating that the pitch angles do not vary with radius. For some galaxies in the stripped sample, the variation between inner and outer disk pitch angles is not much greater than in the control sample, however in a couple of stronger cases, notably JO194, JW39 and JO175, the outer disc pitch angles can be more than double the inner disc pitch angles. It is important to consider that uncertainties in the inclination will affect these numbers. Furthermore, in the case of galaxies which are not fully edge-on stripped, any stripped material which lies extraplanar to the disc will be incorrectly deprojected for inclination, however this should be addressed by taking the average pitch angle at all azimuthal positions. These results indicate that in some cases, the "unwinding" can indeed be characterised by a measurable increase in pitch angle suggesting a loosening of the spiral arms. In other cases, we instead see only "apparent" unwinding, whereby the effect may be the result of the spiral arms being stretched along the direction of stripping, giving the appearance of unwinding without altering the pitch angle.


\begin{table}
\centering
\caption[Comparison of pitch angles in the discs and tails of unwinding and control GASP jellyfish galaxies.]{Comparison of pitch angles in the discs and tails of unwinding GASP jellyfish galaxies. The number of inner and outer pitch angle measurements used in the analysis are listed. Values marked as ? are given where no clear spiral arms are visible.}
\begin{tabular}{l c c c c}\label{table:pitchangles}
\\\textit{Stripped Galaxies}
\\\hline
Galaxy & N$_{\rm inner}$ &$\alpha_{\rm inner} (^\circ )$ & N$_{\rm outer}$ & $\alpha_{\rm outer}(^\circ)$ \\
\hline
JO201	& 0 &?	& 4 & 21.5\\
JO93	& 4 &11.1	& 5 & 21.9\\
JO194	& 2 &13.1	& 2 & 53.4\\
JO200	& 2 &22.0	& 3 & 30.5\\
JO175	& 1 &27.0	& 2 & 66.5\\
JO47	& 0 &?	& 0 & ?\\
JO49	& 2 &26.2	& 2 & 46.5\\
JO70	& 2 &21.3	& 2 & 26.4\\
JO69	& 0 &?	& 0 & ?\\
JO85	& 4 &17.3	& 4 & 23.0\\
JW39	& 1 &7.4	& 1 & 46.3\\
\hline
\end{tabular}

\begin{tabular}{l c c c c}\label{table:pitchangles_control}
\\\textit{Control Galaxies}
\\\hline
Galaxy & N$_{\rm inner}$ &$\alpha_{\rm inner} (^\circ )$ & N$_{\rm outer}$ & $\alpha_{\rm outer}(^\circ)$ \\
\hline
P21734	& 3 &18.8	& 3 & 27.4\\
P25500  & 4 &24.8	& 4 & 18.6\\
P48157	& 3 &19.7	& 2 & 19.4\\
P19482  & 2 &26.9   & 2 & 25.6\\
P63661  & 2 &21.7   & 1 & 21.2\\
P95080  & 2 &14.6   & 2 & 15.6\\
\hline
\end{tabular}
\end{table}


\subsection{Gradients in stellar ages}\label{sec:stellarages}

In order to indirectly estimate the timescale involved in the unwinding effect,
we constrained the spatial distribution of the stellar component, divided into bins of stellar age calculated using the \textsc{SINOPSIS} spectrophotometric fitting code \citep[c.f.][]{Fritz2017}. \textsc{SINOPSIS} finds the best fitting combination of single stellar population spectra to the spectrum of each spaxel, including the pertinent emission lines, in order to estimate the contributions of stellar populations in four age bins 1) $5.7\times 10^9$-$1.4\times10^{10}$yr, 2) $5.7\times10^8$-$5.7\times10^9$yr, 3) $2\times10^7$-$5.7\times10^8$yr and <$2\times10^7$yr to the observed light.

\begin{figure*}
    \begin{tabular}{cccc}
        \subfloat{\includegraphics[width=0.5\columnwidth]{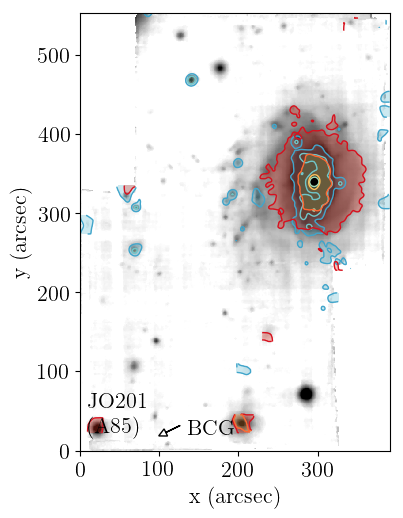}}&
        \subfloat{\includegraphics[width=0.5\columnwidth]{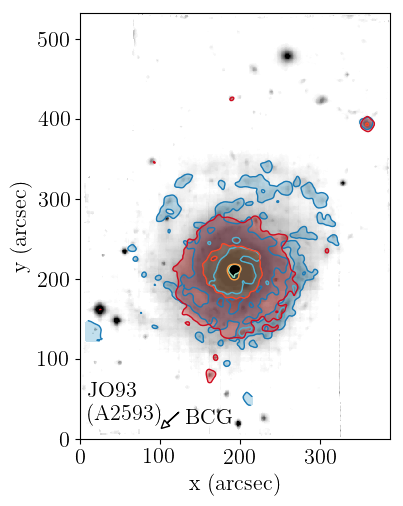}}&
        \subfloat{\includegraphics[width=0.5\columnwidth]{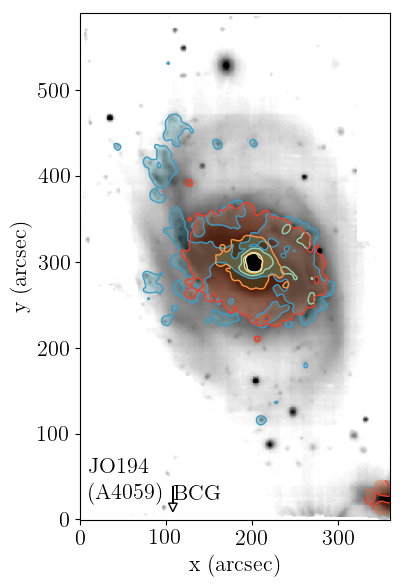}}&
        \subfloat{\includegraphics[width=0.5\columnwidth]{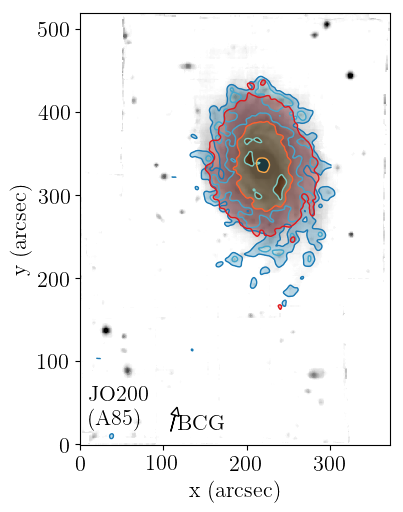}}\\
        \subfloat{\includegraphics[width=0.5\columnwidth]{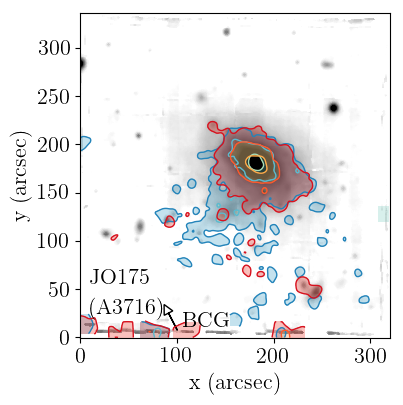}}&
        \subfloat{\includegraphics[width=0.5\columnwidth]{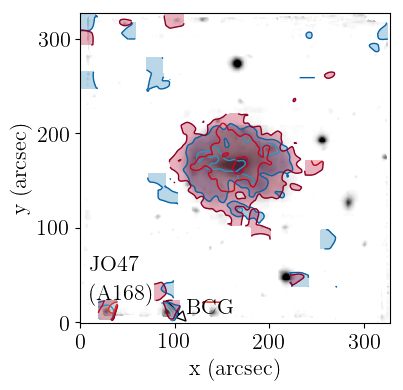}}&
        \subfloat{\includegraphics[width=0.5\columnwidth]{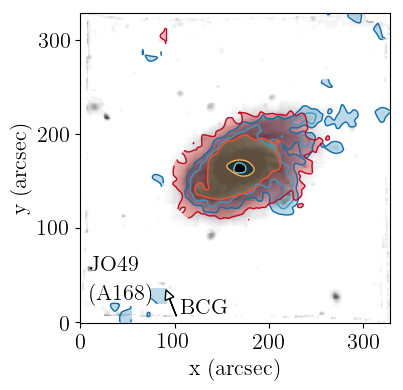}}&
        \subfloat{\includegraphics[width=0.5\columnwidth]{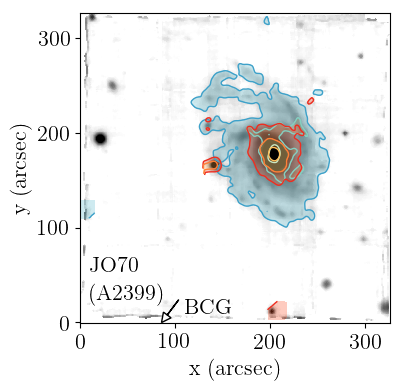}}\\
        \subfloat{\includegraphics[width=0.5\columnwidth]{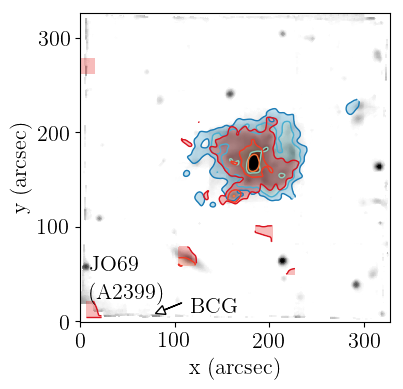}}&
        \subfloat{\includegraphics[width=0.5\columnwidth]{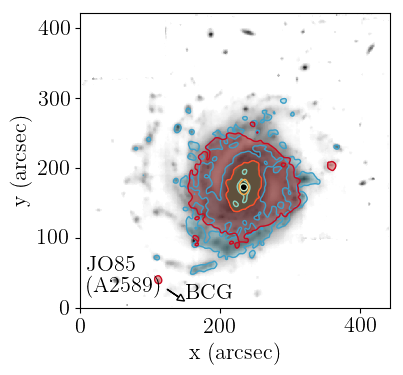}}&
        \subfloat{\includegraphics[width=0.5\columnwidth]{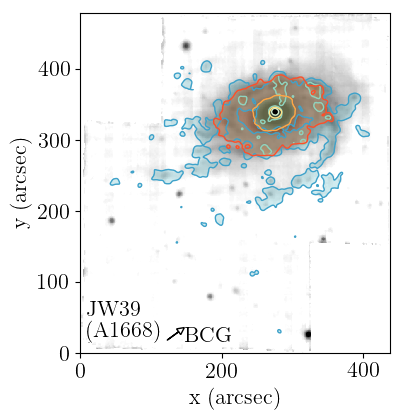}}&
    \end{tabular}
    \caption[Unwinding galaxies shown in white light, overlaid with contours comparing the oldest and youngest stellar populations.]{Unwinding galaxies shown in white light, overlaid with pairs of contours showing the oldest $5.7\times10^9$-$1.4\times10^{10}$yr (red/orange) and youngest <$2\times10^7$yr (blue/light blue) stellar populations calculated using \textsc{sinopsis} spectrophotometric code. The figures show that the older stellar population is generally constrained to the disc of the galaxy and is fairly undisturbed, whilst the younger population of stars extends much further into the tails and describes the extended spiral structure in the unwound arms.}
    \label{fig:stellarpops}
\end{figure*}

Figure \ref{fig:stellarpops} shows the whitelight images of each galaxy, overlaid with the oldest stellar component ($5.7\times10^9$-$1.4\times10^{10}$) in red and the youngest (<$2\times10^7$yr) in blue.
The image reveals that, as expected for purely hydrodynamical disturbance, the older stellar component shown by the red contour line lies solely within each galaxy and appears undisturbed by the stripping. In contrast, the blue contours which trace the newly formed stars more closely follow the pattern of the unwinding spiral arms and extend much further ($\sim20$kpc) outside the galaxy disc. This shows that the stars form out of the spiral arms after they have been stripped and unwound from the galaxy; the newly-formed stars will be unaffected by subsequent stripping.

\begin{figure}
    \centering
    \includegraphics[width=0.43\textwidth]{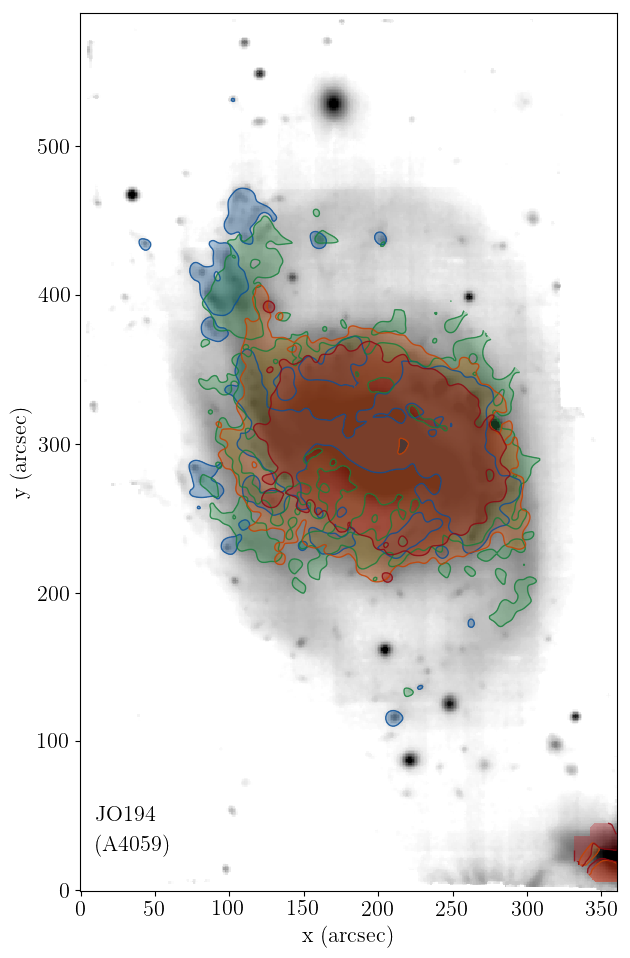}
    \includegraphics[width=0.43\textwidth]{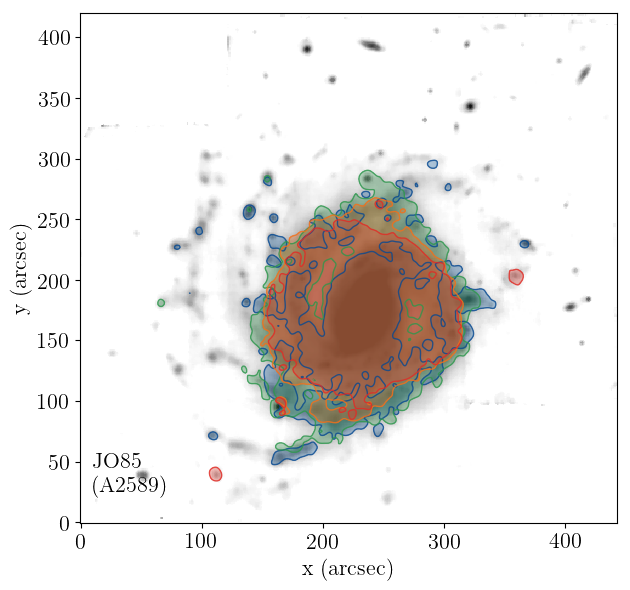}
    \caption[JO194 and JO85 shown in white light, overlaid with four stellar age bins.]{Unwinding galaxies shown in white light, overlaid with four stellar age bins. Red: $5.7\times10^9$-$1.4\times10^{10}$yr, orange: $5.7\times10^8$-$5.7\times10^9$yr, green: $2\times10^7$-$5.7\times10^8$yr and blue: <$2\times10^7$yr.
    
    \textit{Top}, JO194: The upper left arm shows that the youngest (blue) stellar population is much further unwound from the galaxy than the green stellar age bin, indicating that the stars form out of the unwinding gas cloud and remain in the position in which they formed. As the arm unwinds, it has left behind a trail of newly formed stars in its wake.
    
    \textit{Bottom}, JO85: The lower spiral arm shows a similar spatial offset to JO194 between the two youngest stellar populations revealing the unwinding motion of the spiral arm as it leaves a trail of newly formed stars.}
    \label{fig:jo194_ages}
\end{figure}

Highlighting this latter point more precisely, Figure \ref{fig:jo194_ages} shows the most drastic examples of spatial offsets between stellar age bins within the unwinding tails. In both figures, red shows the oldest stellar population ($5.7\times10^9$-$1.4\times10^{10}$yr), followed by orange ($5.7\times10^8$-$5.7\times10^9$yr), green ($2\times10^7$-$5.7\times10^8$yr) with blue showing the youngest (<$2\times10^7$yr). In the upper left of JO194, the green and blue contours show a distinct spatial offset of around 5kpc, which traces the motion of the unwinding arm. The green contour of stars which formed around 300Myr earlier, is more tightly wound with the disc of the galaxy, whilst the blue contour of younger stars lies further extended from the disc, around $6^\circ$ steeper in pitch angle. A similar occurrence is visible in JO85 with the lower spiral arm of the image, with a less prounounced $4^\circ$ increase in pitch angle. This pattern captures the process of the spiral arm opening out and leaving a trail of newly-formed stars in its wake. As the spiral arm material is stripped from the galaxy and begins to undergo star formation, the stars, as mentioned before, remain unaffected by subsequent ram-pressure, resulting in an age-gradient. In many ram-pressure stripped galaxies, such age-gradients are observed along the knots of star formation \citep{Sheen2017,Jachym2019} as diffuse gas collapses along its motion trailing behind the galaxy  \citep{Yoshida2008,Kenney2014}, here the same effect reveals the outstretching of the spiral arm during the stripping process.


\section{Simulations}\label{sec:sims}


To understand more precisely the process, or processes, by which hydrodynamical interactions can result in unwound spiral arms, we compare the observed morphology and velocity field of our sample to a set of high resolution wind-tunnel simulations of an intermediate mass spiral galaxy undergoing ram pressure stripping. Previous works have shown that the inclination of the ICM wind to the galaxy disk affects the efficiency of stripping \citep{Roediger2006,Jachym2009}, and have shown that unwinding can occur in certain circumstances \citep{Schulz2001,Roediger2014,Steinhauser2016}. In this study, we comprehensively analyse the evolution of the unwinding pattern over time for galaxies at different inclinations, comparing directly with the morphology and kinematics of the observed sample and features detected with MUSE. The simulations consider fixed inclination angles for the direction of the ram pressure wind, face-on and edge-on with respect to the plane of the model's disc. In this way, we hope to better understand the role of inclination on the velocity field of the gas, as well as the effectiveness with which the spiral arms are unwound and the pattern of unwinding.

The model spiral galaxy we use has three main components of mass - an NFW dark matter halo, exponential disc of stars, and exponential disc of gas. Our model galaxy does not contain a bulge component. The dark matter halo mass of the galaxy is $\mathrm{M}_\mathrm{total}=3.2\times 10^{11}\mathrm{M}_\odot$, with a virial radius of 140~kpc, and halo concentration=12, consisting of 5 million dark matter particles. The stellar disc has a total mass of  $\mathrm{M}_*=1\times 10^{10}\mathrm{M}_\odot$, with a radial exponential scalelength of 2.2~kpc, truncated at 4 scalelengths, initially consisting of 250 thousand star particles. The gas disc initially has a total mass of $\mathrm{M}_\mathrm{gas}=1.3\times 10^{9}\mathrm{M}_\odot$, and a radial scalelength of 3.7~kpc, truncated at 4 scalelengths, and is assumed solar metallicity. Initial conditions were built using the publically available initial-conditions set-up code {\sc{DICE}} \citep{Perret2014}. Tests demonstrate that the initial conditions were very stable, and were evolved for 0.5~Gyr in isolation to ensure dynamical stability had been reached.

Simulations were carried out with the Adaptive Mesh Refinement (AMR) code {\sc{RAMSES}} \citep{Teyssier2002}. The total volume was a box with side-length of 280~kpc, that fully encloses the virial radius of the dark matter halo. The adaptive mesh was allowed to refine according to the mass within a cell to a maximum refinement level=14, equivalent to a smallest cell-size of 17~pc. In addition, the refinement level ensures that at least four Jeans-lengths are resolved, down to the maximum level of refinement. The simulation considers a radiative-cooling treatment for solar metallicity gas. Star formation occurs following a standard prescription of a Schmidt relation with a 3\% star-formation efficiency for all gas above a critical density threshold of $0.1\mathrm{cm}^{-3}$. Supernova feedback is treated thermally, with an efficiency of 20\%, and a yield of 0.1. We confirm that, in isolation, our model disc lies on the star-forming main sequence.

Ram pressure was modelled using a wind-tunnel simulation set-up. A hot, fast flowing, low density, intracluster medium gas is fed into the simulation box from one wall of the simulation box. We initialise the disc with a low constant wind to roughly imitate the outskirts of the cluster. Then, after reaching 1~Mpc from the cluster centre, the density and inflow velocity of the intracluster medium begins to smoothly evolve with time, to mimic the changing ram pressure that a cluster spiral might experience when falling into a real cluster. To calculate the time evolving density, we assume a beta model for the intracluster medium that is a rough approximation for the Virgo cluster (central density=$2.0 \times 10^{-26}$~g/cm$^3$, $\beta=0.5$, core radius=50~kpc). To calculate the time evolving wind speed, we calculate the orbital velocity of a galaxy moving through the cluster potential well, and assume the inflow speed matches the orbital velocity. Under the assumption of hydro-static equilibrium with a gas temperature of $4.7 \times 10^7$~K, the potential well of the cluster is fully defined. Using a simple time-step integrator, we trace out the orbit of a galaxy falling in from 1~Mpc. We assume the galaxy has an initial radial velocity component of 565~km/s, and a tangential velocity component of 565 km/s. This results in quite a plunging orbit reaching a pericentric distance of 340~kpc from the cluster centre. During the infall, the orbital velocity increases from 800~km/s and reaches a maximum velocity of $\sim$1600~km/s at pericentre, and we set the inflow speed of the gas in our wind-tunnel test as equal to the orbital velocity.

For each of the inclination angles considered, we conduct the simulation for at least 2.3~Gyr in total. This is sufficient to pass the cluster pericentre, which occurs after approximately 1.7 Gyr. For this study, we find that the heavily truncated gas discs of those galaxies that have passed pericentre do not show features resembling the unwrapping we see in our observational sample.
Therefore, we restrict our attention to the period between when the model discs enter the cluster and just before they reach pericentre.

Now we create "observed" velocity maps of the simulated discs at various instants in time, as they approach the cluster pericentre for the first time. We consider instants when the galaxy is at a cluster centric radius of $r_{\rm{cl}}$=0.92~Mpc (in the cluster outskirts), 0.66~Mpc (intermediate radius), and 0.35~Mpc (just before pericentre). At each of these instants we produce maps by binning the gas properties along a given line of sight, considering similar resolution as in the MUSE observations. We produce maps for both disc inclinations. As we are comparing to $\mathrm{H}\alpha$ emission-line maps, we only consider gas above the star formation threshold in the simulation (i.e., with density $>0.1\mathrm{cm}^{-3}$) in our maps. To produce kinematics maps for comparison with the GASP observed galaxies, the data were transformed to different lines-of-sight and binned across the plane of observation to a similar resolution to MUSE, then the mean velocity within each bin was taken to produce the velocity maps.


\subsection{Comparing Observations with Simulations}\label{sec:kin_compare}

JO200 and JO85 were selected as suitable galaxies for comparison with the simulation models, given the clarity of their disc and spiral arms, as well as strongly symmetric and asymmetric unwinding respectively, as definitive cases for each. The masses of the observed galaxies do not match the masses of the simulated examples, but we are only concerned here about the pattern produced by the interaction.

\begin{figure*}
    \centering
    Observed\hspace{235pt}Simulated\hspace{10pt}\hphantom{a}
    \includegraphics[width=\textwidth]{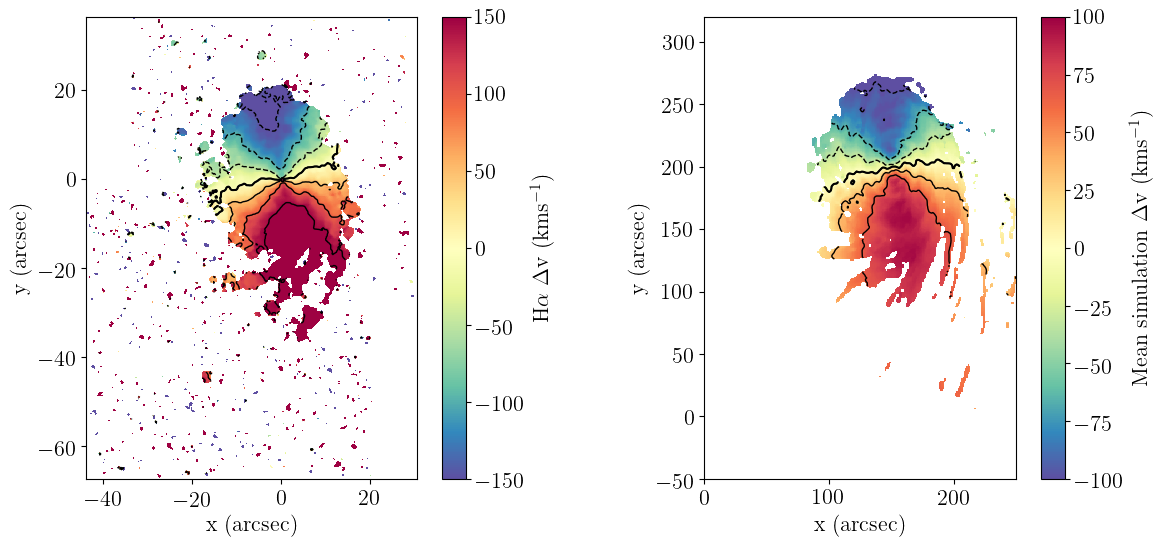}
    \includegraphics[width=\textwidth]{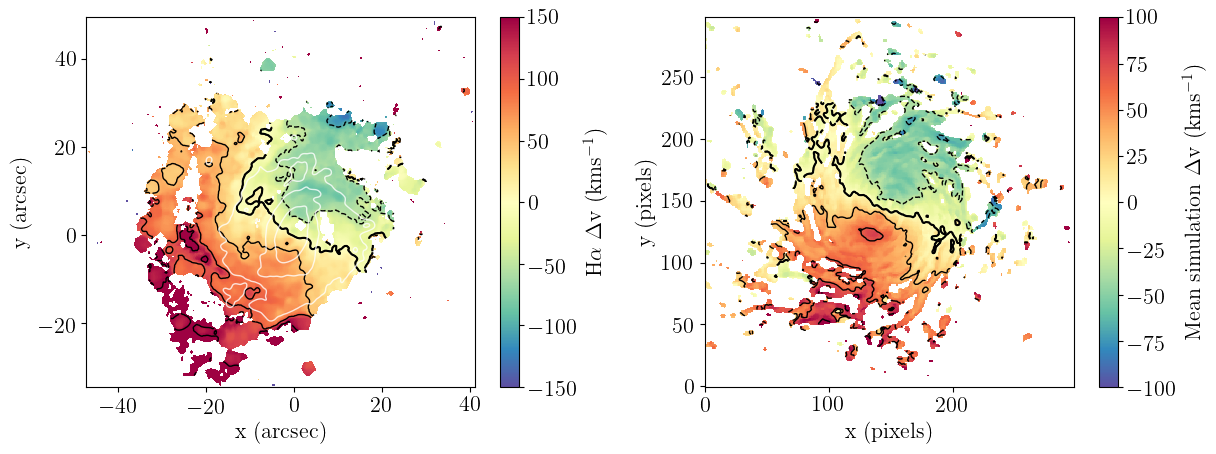}
    \caption[Kinematics comparison between JO200 and a simulated galaxy undergoing edge-on stripping.]{\textit{Top Left:} Kinematics map of JO200 measured using H$\alpha$ emission line velocities from \textsc{kubeviz}. \textit{Top Right:} Kinematics map of "observed" simulated galaxy undergoing edge-on stripping, produced as described in section~\ref{sec:kin_compare}. The simulated unwinding galaxy is rotated to match the inclination and position angle of JO200, and matches the unwinding pattern and kinematics fairly well, suggesting that JO200 is well-described by an edge-on stripping scenario.
    \textit{Bottom Left:} Kinematics map of JO85 measured using H$\alpha$ emission line velocities from \textsc{kubeviz}. \textit{Bottom Right:} Kinematics map of "observed" simulated galaxy undergoing face-on stripping. The simulated unwinding galaxy is rotated to match the inclination and position angle of JO85. The overall similar kinematics suggest that JO85 may be undergoing close-to edge-on stripping, but differences and asymmetries suggest that some component of edge-on motion is also present.}
    \label{fig:kin_compare}
\end{figure*}

The compared velocity fields of JO200 and the edge-on simulated galaxy are shown in Figure~\ref{fig:kin_compare}. For ease of comparison between the two, the simulated galaxy was rotated by transforming the data to match the inclination and position angle of JO200, mimicking the observed line of sight of the galaxy. The mean particle velocities along each pixel "bin" gave the kinematics maps shown in the figure.

Whilst the absolute values of the rotation curve differ as a result of the different total masses of the real and simulated galaxies (JO200 $\mathrm{M}_{*}=7\times10^{10}$, simulation $\mathrm{M}_{*}=1\times10^{10}$), the gradient of the rotation is very similar in both cases. It is apparent from the figure that the velocity contours in the observed galaxy are more "U"-shaped at the most extreme ends of the rotation curve, while the simulated galaxy exhibits a more uniform rotation. The spiral arms in both the simulated and observed galaxies show a similar amount of unwinding with similar pitch angles (JO200: inner=22.0 outer=30.5, Edge-on sim: inner=15.6 outer=30.5) in the unwound tails at the lower edge of the galaxies. The resemblance suggests that the unwinding pattern of JO200 may be well described by a nearly edge-on stripping scenario.

The velocity pattern in JO85 shows similarities to the face-on stripping simulated galaxy, particularly in the curvature of the zero contour where the top end bends to the right as the stripping begins to skew the velocities at the edge of the disc. The similarities suggest that the JO85 is experiencing close to face-on motion, but differences and asymmetries suggest that some component of edge-on motion is also present. The kinematic patterns which evidence face-on and edge-on interactions will be further explored in the simulations in the next section.

\subsection{Tracing Unwinding Throughout Infall}\label{sec:sim_timesteps}

\begin{figure*}
    \centering
    \includegraphics[width=\textwidth]{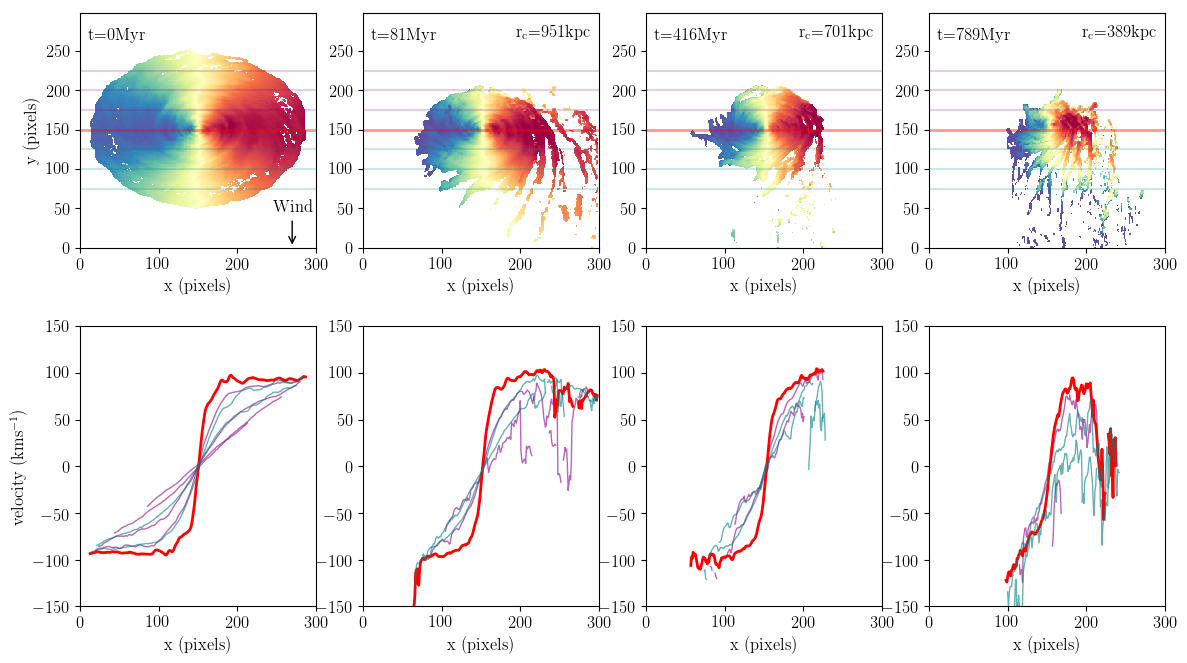}
    \caption[Timestep snapshots of a simulated galaxy undergoing edge-on stripping, moving partially along the line-of-sight.]{Timestep snapshots of a simulated galaxy undergoing edge-on stripping, inclined $45^\circ$ with respect to the observer such that the velocity map can be extracted to emulate a physical observation. In this example, the motion of the galaxy is 50\% along the line-of-sight away from the observer, and 50\% upward along the plane of the observation. The wind arrow indicates the projected wind direction on the plane of the image; an additional, equal component is toward the observer. The snapshots are labelled with the time since entering the cluster on the upper left, as well as the clustercentric distance in the upper right for each of the infall timesteps. In the lower panels, the red lines show the central rotation curves, whilst the lighter magenta and cyan lines show the rotation curves across the top and bottom of the disc of the galaxy respectively, corresponding to the lines on the upper panels.}
    \label{fig:sim_perp}
\end{figure*}

\begin{figure*}
    \centering
    \includegraphics[width=\textwidth]{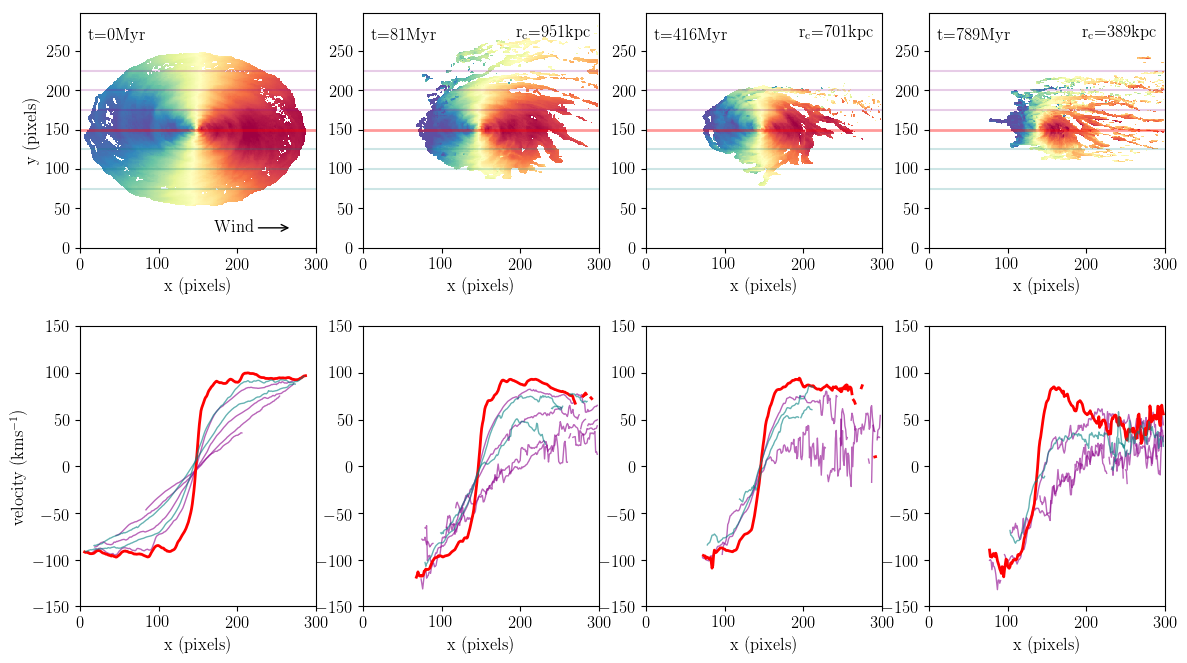}
    \caption[Timestep snapshots of a simulated galaxy undergoing edge-on stripping, moving along the plane of the sky.]{Timestep snapshots of a simulated galaxy undergoing edge-on stripping, inclined $45^\circ$ with respect to the observer such that the velocity map can be extracted to emulate a physical observation as for Figure~\ref{fig:sim_perp}. In this case, the motion of the galaxy is entirely across the plane of the sky, towards the left of the page. The wind arrow indicates the wind direction, which in this case is entirely along the arrow vector, on the plane of the page. The snapshots are labelled with the time since entering the cluster on the upper left, as well as the clustercentric distance in the upper right for each of the infall timesteps. In the lower panels, the red lines show the central rotation curves, whilst the lighter magenta and cyan lines show the rotation curves across the top and bottom of the disc of the galaxy respectively, corresponding to the lines on the upper panels.}
    \label{fig:sim_par}
\end{figure*}

We further studied the mechanisms behind the process of unwinding by analysing the kinematics and shape of the simulated galaxies over several timesteps throughout the infall process.
Snapshots were taken of the edge-on and face-on galaxies during three stages (81, 416, 789Myr) throughout the infall process. Additionally a snapshot taken prior to infall is shown for the edge-on galaxy. These snapshots were binned to a grid of pixels at a similar resolution to the observations and, in the case of edge on, analysed in two "observed" orientations: 1) the galaxy is falling partially along the line of sight, and 2) the galaxy is falling across the plane of the sky. The snapshots of the two cases are shown in figures \ref{fig:sim_perp} and \ref{fig:sim_par} respectively.

In the line-of-sight case, figure~\ref{fig:sim_perp}, it can be seen that the gas on the advancing edge of the galaxy, the right hand edge in the figure, is slowed by the ram-pressure as it rotates into the wind. As the wind slows the gas, it causes it to build up as well as fall to higher orbits, producing a large asymmetry in the second timestep. As the simulation progresses, the stripped gas falls further behind the galaxy and the tail becomes more symmetric. In the final timestep, after the outer layers of gas have been removed, the remaining object is left with a very steep rotation curve gradient.

In the plane-of-the-sky case, figure~\ref{fig:sim_par}, a similar asymmetry is observed in the shape of the disc in the second timestep, arising from the gas being slowed by the ICM wind, causing it to fall to higher orbits. This asymmetry shortly vanishes and the tail becomes more collimated by the final timestep. The rotation curves show the velocity curve being "dragged" along the wind direction to the left and the extreme values of the rotation curve decrease over time. As with the galaxy moving mostly along the line of sight, the velocities in the central region are mostly unaffected and the remaining object retains a steep rotation curve.

\begin{figure*}
    \centering
    \includegraphics[width=\textwidth]{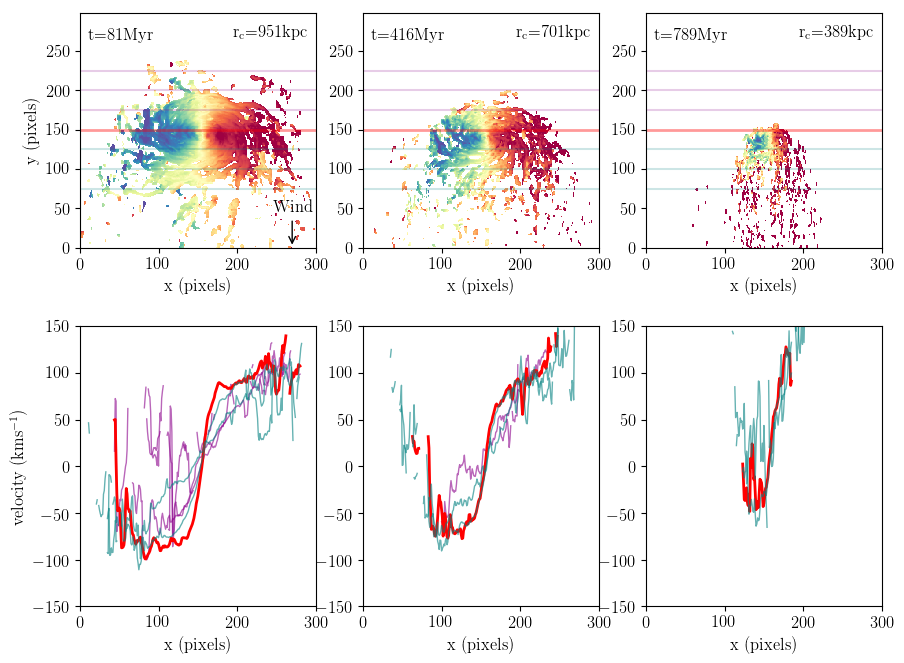}
    \caption[Timestep snapshots of a simulated galaxy undergoing face-on stripping.]{
    Timestep snapshots of a simulated galaxy undergoing face-on stripping, inclined $45^\circ$ with respect to the observer such that the velocity map can be extracted to emulate a physical observation. In this example, the motion of the galaxy is 50\% along the line-of-sight toward the observer, and 50\% upward along the plane of the observation. The wind arrow indicates the projected wind direction on the plane of the image; an additional, equal component is away from the observer. The snapshots are labelled with the time since entering the cluster on the upper left, as well as the clustercentric distance in the upper right. In the lower panels, the red lines show the central rotation curves, whilst the lighter magenta and cyan lines show the rotation curves across the top and bottom of the disc of the galaxy respectively, corresponding to the lines on the upper panels.}
    \label{fig:sim_fo}
\end{figure*}

Figure~\ref{fig:sim_fo} shows the simulated face-on infalling galaxy in 3 timesteps starting from initial infall.
The ICM wind direction is downward and slightly inclined away from the observer. The velocity curve of the galaxy shows a disturbed but still fairly symmetric rotation in the initial timestep, with a slight deflection at the edges to higher velocities, similar to that observed in JO201 \citep[c.f. Figure~15 within][]{Bellhouse2017}.
In the second timestep, more of the gas has been pushed to higher velocities and the central rotation curve is noticeably disturbed. By the third timestep, most of the gas has been pushed by stripping to higher velocities and the remaining object retains a steep rotation curve.

These simulations highlight three examples of the effects of orientation of an infalling galaxy with respect to both the ICM wind and the observer, revealing the differences in the effect of face-on and edge-on stripping on the velocity profile. In edge-on stripping scenarios, the differential velocities between the edges of the galaxy rotating into and out of the wind appears to result in a difference in the effectiveness of stripping, such that the gas rotating with the wind direction is more readily stripped, whilst the gas rotating against the wind is slowed initially, causing an asymmetric build-up of material on the corresponding trailing side.

In the face-on stripping scenario, the gas is stripped around the edges of the disc, causing the rotation curve across all areas of the disc to be much more disturbed, upward deflections on either side of the rotation curve indicate that the stripping occurs evenly around the disc, in contrast with the edge-on scenario. The resulting galaxy after the face-on stripping has a shorter, steeper rotation curve and is fairly compact, likely to be owing to the increased effectiveness of face-on stripping \citep{1999MNRAS.308..947A,2000Sci...288.1617Q,2001ApJ...561..708V, Roediger2006, Jachym2009} removing gas from the disc down to a smaller truncation radius.
The differences between the velocity maps in these scenarios suggest that during the phase of unwinding, the morphology of the spiral arm pattern and its underlying kinematics can be a probe of the inclination of a galaxy's stripped gas, and therefore motion, with respect to the ICM.

In all of the simulated galaxies, the final snapshots of figures \ref{fig:sim_perp}, \ref{fig:sim_par} and \ref{fig:sim_fo} show an extended tail featuring the tentacles and knots commonly observed in stripped galaxies. This suggests that as the residual rotation is slowed and the ICM wind fully washes out the unwinding pattern, the trails of denser gas within the unwound spiral arms become distributed into the tentacles and trails of star-forming knots associated with jellyfish galaxies.

\begin{figure*}
    \centering
    \includegraphics[width=\textwidth]{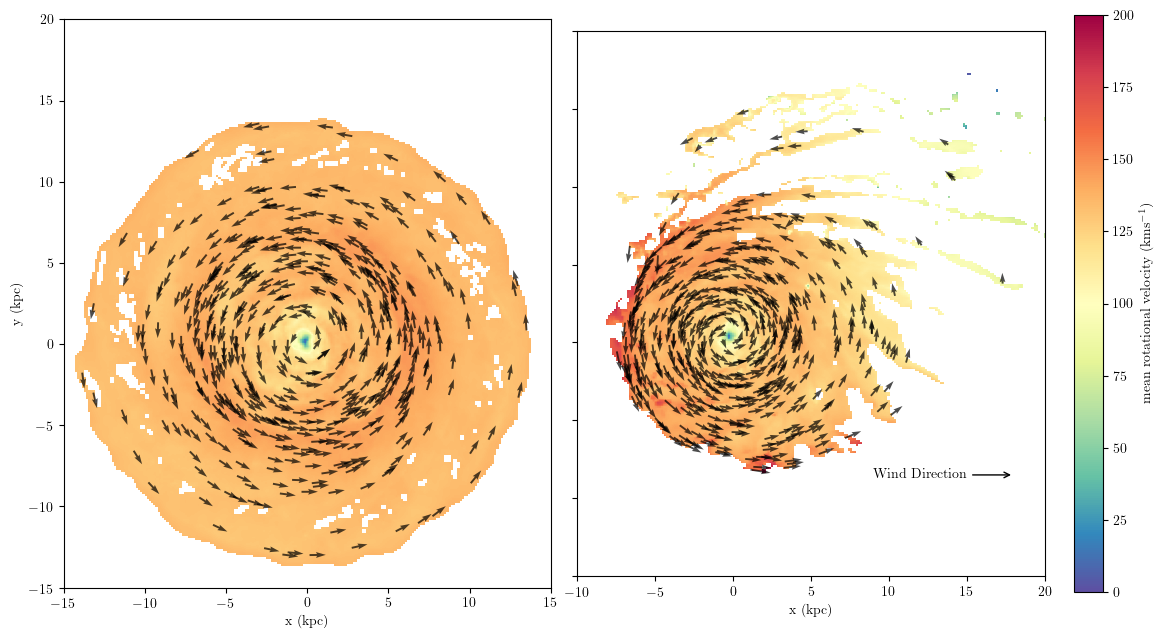}
    \caption[]{Flow diagram showing the 2D velocities of gas particles in the isolated (left) and edge-on stripping(right) galaxies. Colours show the mean 2D velocity in each pixel bin, whilst arrows indicate the 2D velocity vector of 500 randomly selected particles.}
    \label{fig:flow_eo}
\end{figure*}

\begin{figure*}
    \centering
    \includegraphics[width=\textwidth]{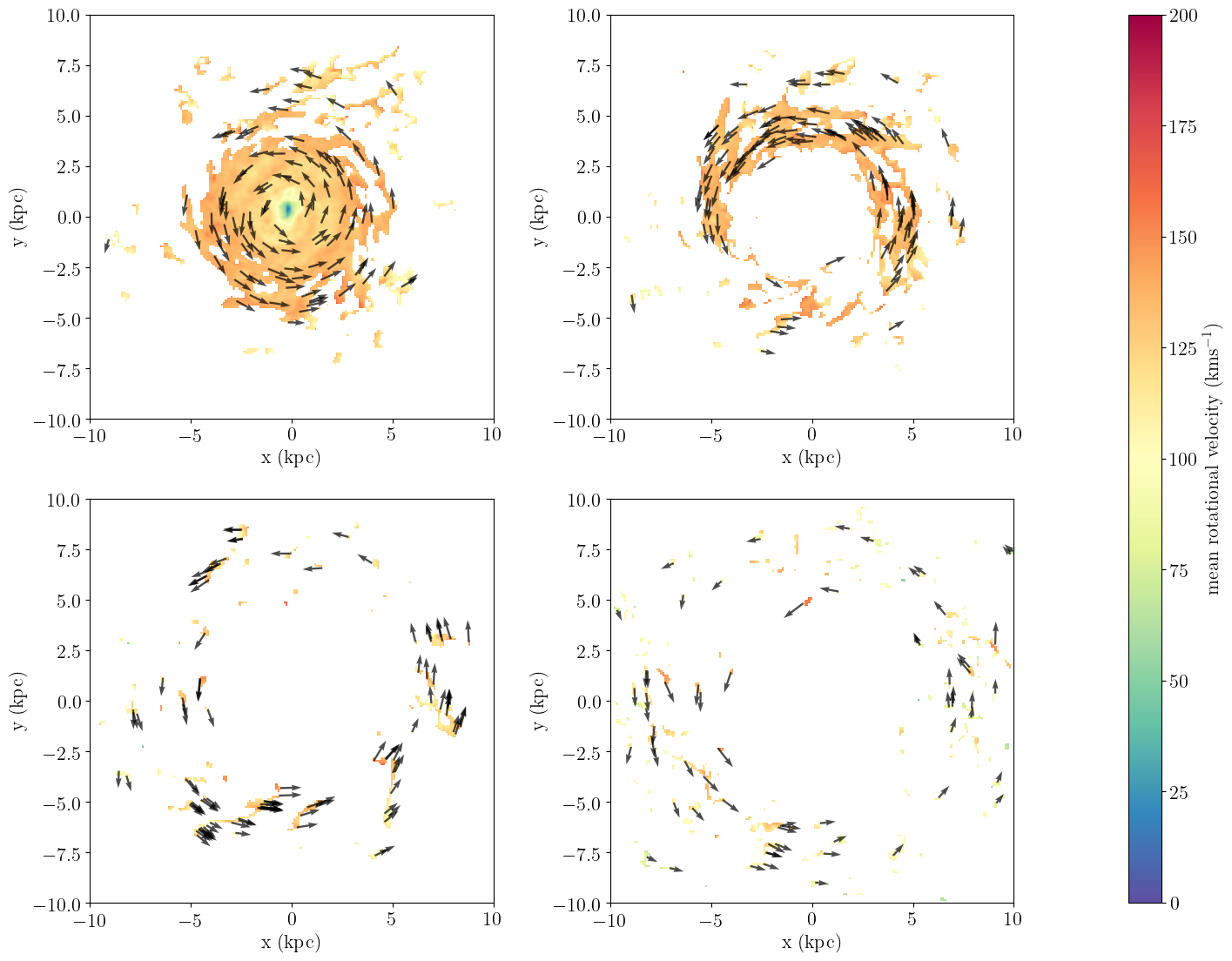}
    \caption[]{Flow diagram showing the 2D velocities of gas particles in four slices of the galaxy undergoing face-on stripping, taken at different distances from the plane of the galaxy. Top left: <1kpc, top right: 1-2kpc, bottom left: 2-3kpc, bottom right: >3kpc. Colours show the mean 2D velocity in each pixel bin, whilst arrows indicate the 2D velocity vector of 100 randomly selected particles in each slice.}
    \label{fig:flow_fo}
\end{figure*}

Highlighting the different "unwinding" processes resulting from edge-on and face-on stripping, 2D flow diagrams were produced to show the motion of the rotating gas. The diagrams were produced by measuring the mean 2D velocity in bins across the extent of the plane of the disc of each galaxy. The resulting diagrams, shown in figures~\ref{fig:flow_eo} and \ref{fig:flow_fo}, show the mean scalar 2D velocity in the colourmap, along with arrows indicating the 2D velocity vectors for a subsample of particles in each snapshot. In Figure~\ref{fig:flow_eo} the isolated, undisturbed galaxy is shown alongside the edge-on stripped galaxy. The mean 2D velocities in each pixel are calculated for all particles within that pixel bin along the line of sight and the vectors are shown for 500 randomly selected particles in each snapshot. In Figure~\ref{fig:flow_fo}, showing the face-on stripping galaxies, the flow diagrams are shown for a single snapshot broken into plane-parallel slices along the direction of stripping at intervals of 1kpc. The arrows are drawn for 100 randomly-selected particles in each slice.
The edge-on stripped galaxy flow diagram further evidences the differential ram-pressure caused by the rotation of the galaxy. The gas rotating with the wind, at the bottom edge of the galaxy in the figure, is sped up initially and any gas which remains bound to the galaxy falls to higher orbits conserving angular momentum. The gas rotating into the wind is not completely stripped, but slowed by the ram-pressure, causing a "pile-up" of material as the gas continues to rotate. This has the observed effect of the spiral arms opening out and extending on the edge of the galaxy rotating against the wind.
Interestingly, the rotational velocity on the leading edge of the disc increases in comparison to the isolated galaxy. This is likely to be due to the gas being pushed deeper into the disc by ram-pressure, conserving its angular momentum and increasing its rotational velocity.
Furthermore, this figure highlights an important feature of the initial morphology of the tails with respect to the galaxy. In this initial timestep, within the first 81Myr of stripping, the majority of the "tail" lies at $45^\circ$ to the wind direction due to the ongoing rotation of the stripped material. The majority of observed jellyfish galaxies are seen at more advanced stages of stripping and thus, in projection, the tail directions can be assumed to closely match the wind direction. For galaxies in early stages of infall, however, particularly those with strong edge-on "unwinding" morphologies, care should be taken when utilising tail directions to estimate the projected direction of motion.

In Figure~\ref{fig:flow_fo} the four panels show slices taken along the direction of stripping, at 1kpc intervals, for the face-on stripped galaxy. The flow diagram shows the rotation of the gas within <1kpc of the galaxy (top left), 1-2kpc (top right), 2-3kpc (bottom left) and >3kpc (bottom right) from the plane of the disc. The figure clearly shows the expanding ring of stripped material which, once dislodged from the edges of the disc, falls to higher orbits as it becomes further removed from the potential well of the galaxy. This manifests itself in the spiral arm pattern as an extending of the spiral arms, which loosen and unwind as the rotation slows.

\section{Summary and Discussion}\label{sec:discussion}



We find that a pattern of unwinding spiral arms is visible in 11 face-on galaxies in the GASP sample which are undergoing ram-pressure stripping interactions during cluster infall.

The sample of galaxies was selected visually to include galaxies viewed mostly face-on, with clear spiral arms. Out of 12 galaxies which satisfied these first criteria, 11 showed possible signs of unwinding spiral arms. The high incidence of the unwinding effect among face-on viewed galaxies with visible spiral arms in the GASP sample suggests that this effect could be particularly common, although this may be dependent on the original visual sample selection of GASP.

Analysis of the morphology in different wavelengths revealed that this pattern appears only in the stripped $\mathrm{H}\alpha$ emitting ionised gas, and consists of extended, curved trails of material which continue along the lines of the spiral arm patterns. Furthermore, the extent of unwinding around the edges of the disk may indicate the inclination of stripping with respect to the wind, as the more face-on encounters appear to show unwinding around a greater extent of the disk. The morphology and kinematics both reveal that the stripping and unwinding do not affect the older stellar component and the inner discs of the galaxies appear largely undisturbed, confirming that the effect is resulting from purely hydrodynamical processes. Furthermore, with the possible exception of JO69, none of the galaxies in the sample have sufficiently close companions that could plausibly induce a gravitational disturbance.

The radial variations of the pitch angles were compared in each of the stripping and undisturbed galaxies, as well as the simulated stripped galaxies. The sample of stripped galaxies were found to have generally higher pitch angles in the outer spiral arms compared to those within 2 effective radii, in contrast to the 6 control galaxies which have generally constant pitch angle throughout the disc. This result was also confirmed in the simulated stripped galaxies, which showed similar increases in pitch angle to the observed sample. This confirms that the visual unwinding effect is indeed a pattern in the spiral arms "opening" in the outermost stripped regions of the galaxy. In a few examples, the pitch angle was found to vary negligibly, which suggests that for those cases, the lengthening of the spiral arms resulting from stripping can also give an appearance of unwinding without actually affect the pitch angles. \citet{Schulz2001} noted possible winding of the inner spiral arms due to compression and "annealing" of the disc in addition to unwinding of the outer regions of their face-on stripped galaxies. With the given sample we do not have sufficient evidence to investigate this, however with a larger sample of galaxies the effect of ram-pressure stripping on the more central spiral arms could be tested.

The unwinding tails were found to be comprised of only the youngest stellar populations, in many cases only stars <20Myr, when analysed with \textsc{sinopsis} spectrophotometric fitting code. These have likely formed in-situ, confirming that the unwinding effect only occurs in the gas component of the galaxy. Moreover the age gradients observed in some unwinding arms possible capture the process in action, revealing a trail of progressively younger stars left by the opening arms of stripped material.

Simulations were carried out of ram-pressure stripping interactions in different inclinations with respect to the ICM wind to investigate how the unwinding effect can arise and to compare the resulting morphology with observed features in the MUSE sample. The resulting snapshots were analysed throughout the stripping process, and used to produce "observations" from different viewing angles to compare with the observed sample of galaxies.

The two main effects which can, individually and in combination, give rise to an unwinding pattern are as follows:
\begin{enumerate}
    \item In edge-on wind interaction cases, the differential ram-pressure in the leading and trailing edges of the disc causes a build-up of material in the leading edge. The tail becomes asymmetric and the slowed rotation of the gas causes the pitch angle to increase.
    \item In face-on wind interactions, the gas from the outer spiral arms is stripped and withdrawn from the potential well of the galaxy disc, causing it to move to higher orbits and increase the pitch angle in all directions from the disc.
\end{enumerate}
In most cases, a combination of these two effects will occur as a galaxy is stripped at angles between face-on and edge-on with respect to the ICM wind. Cases in which the galaxy is moving mostly edge-on to the wind appear to be more asymmetric and have tails that are further collimated to one side of the disc, whilst cases closer to face-on such as JO201 and JO85 show gas extending in a much wider range of directions from the disc.

Furthermore the simulations, combined with the phase-space positions of our galaxies and the \textsc{sinopsis} stellar population analysis, as well as similar effects found in the literature \citep{Steinhauser2016,Roediger2014} all point toward the idea that the unwinding effect occurs only for galaxies undergoing first infall on radial plunging orbits, and appears to be a short-lived phase during stripping, containing stars formed in-situ no more than 20Myr ago in our observations and appearing to be fully developed by 81Myr after first infall in the simulations. After this phase, the simulations suggest that the unwound arm features subsequently evolve over the next $\sim700$ Myr into the tentacles and trailing knots which are ubiquitous in many other jellyfish galaxies.

It is important to stress that, whilst we show here that a purely hydrodynamical process can give rise to a pattern of unwinding in the spiral morphology of a galaxy, this does not imply that the unwinding effect is a "smoking gun" of ram-pressure stripping. It is certainly the case that in an observed unwinding galaxy, effects such as tidal interactions and minor mergers may also be responsible. What we have found here is that simulated galaxies experiencing pure hydrodynamical interactions, in the absence of gravitational effects, can successfully reproduce the morphological and kinematic patterns found in an observed sample of galaxies. It is also important to highlight that the unwinding pattern is not always the sole mark of the interaction and is often compounded with other effects such as compression of the disc and the formation of extended tails and filaments in the wake of the galaxy's motion. The simulations presented here indicate that the unwinding pattern is mostly present in the early stages of stripping, whilst other effects develop thereafter. The unwinding pattern may therefore be dominant in galaxies initially undergoing stripping and will become washed out as other indicators of stripping gain prevalence, as the galaxy proceeds through infall.

The results of the simulations and the observed differences between patterns in the tails in different regions of phase space suggest that a relation may exist between the stripped morphology and the inclination of the galaxy with respect to the ICM wind, such that broader tails may be indicative of stripping at angles close to face-on to the wind and, vice versa, narrow tails may result from stripping closer to edge-on, as outlined by the two simulated examples in Figure~\ref{fig:kin_compare}. With a sufficiently large sample, it may be useful to further explore the possibility of such a relation.

An important key result of this study is that we confirm through simulations and observations that hydrodynamical interactions alone can produce an unwinding effect in spiral galaxies, consisting of curved tails and arcs of stripped material resulting from the residual rotation of the stripped material; ram-pressure stripping can produce not only straight tentacles, but also arced tentacles under the right conditions. Since curved tails and stripped arcs can be indicative of tidal interactions, care should be taken to confirm the origin of stripping when classifying samples of interacting galaxies based on these particular morphological features. This is particularly important when considering selection of ram-pressure stripped galaxies in broad-band imaging data by manual visual inspection or automated feature-detection; galaxies which might otherwise be dismissed due to "tidal-like" features should be carefully checked. In our sample, we noted that all of the galaxies exhibited some other signatures of stripping, however the simulations revealed that certain orientations can give rise to unwinding with no other obvious effects, which could plausibly be interpreted as evidence of a gravitational, not hydrodynamical, interaction. In cases where no obvious companion is present to confirm gravitational interactions, observations of the stellar and gas rotation curves are likely the best indicators of the true cause of morphological disturbance.

\section*{Acknowledgements}

Based on observations collected at the European Organization for Astronomical Research in the Southern Hemisphere under ESO programme 196.B-0578. This project has received funding from the European Research Council (ERC) under the European Union's Horizon 2020 research and innovation programme (grant agreement No. 833824).

C.B. would like to thank Frederic Vogt for valuable discussion and guidance in developing the azimuthal reprojection plots, based on figures shown in \citet{Vogt2017}.

Y.J. acknowledges financial support from CONICYT PAI (Concurso Nacional de Inserci\'on en la Academia 2017) No. 79170132 and FONDECYT Iniciaci\'on 2018 No. 11180558.

We acknowledge financial contribution from the contract ASI-INAF n.2017-14-H.0, from the grant PRIN MIUR 2017 n.20173ML3WW\_001 (PI Cimatti) and from the INAF main-stream funding programme (PI Vulcani).


    

\section*{Data Availability}
The data underlying this article are available in http://archive.eso.org/scienceportal/home under programme ID 196.B-0578.

\bibliographystyle{mnras}
\bibliography{references.bib}

\appendix

\section{Kinematics Comparisons of Galaxy Sample}

This Appendix follows on from figure~\ref{fig:kin_compare_vels} with the velocity map comparisons for the rest of the sample of unwinding galaxies, with left panels showing the H$\alpha$ emitting ionised gas velocity and right panels showing the stellar velocities, revealing that the majority of the disturbance is visible only in the stripped gas and not in the old stellar component. This is indicative that the disturbance is hydrodynamical in nature and not resulting from gravitational effects.

\begin{figure*}
    \centering
    \includegraphics[width=\textwidth]{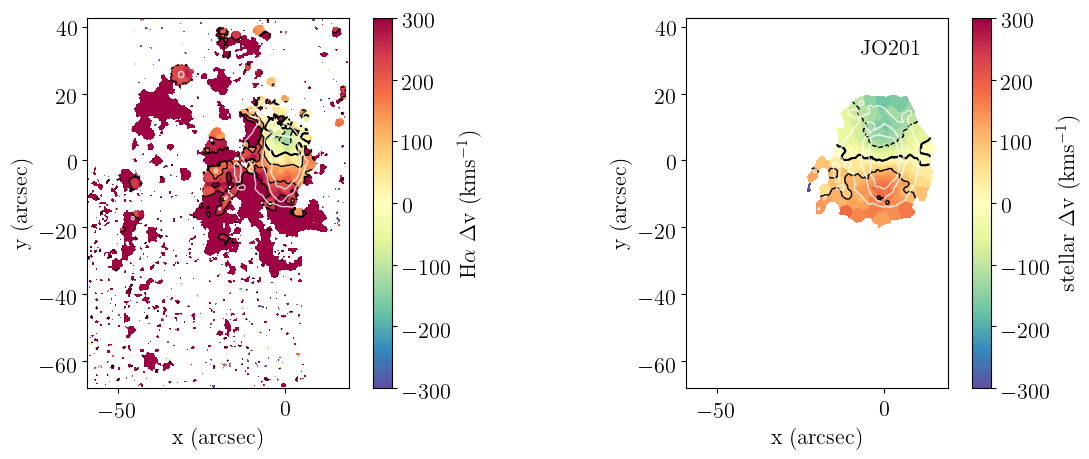}
    JO201
    \includegraphics[width=\textwidth]{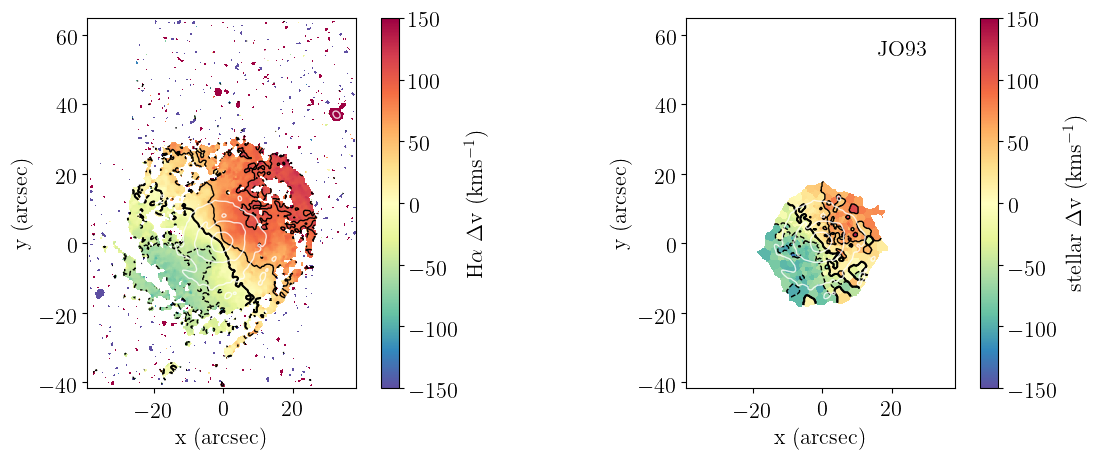}
    JO93
    \includegraphics[width=\textwidth]{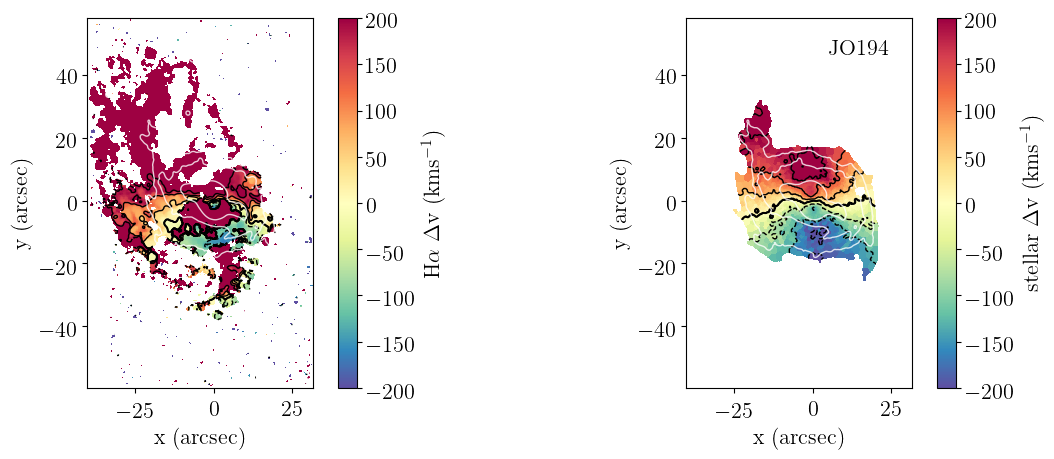}
    JO194
\end{figure*}
\begin{figure*}
    \centering
    \includegraphics[width=\textwidth]{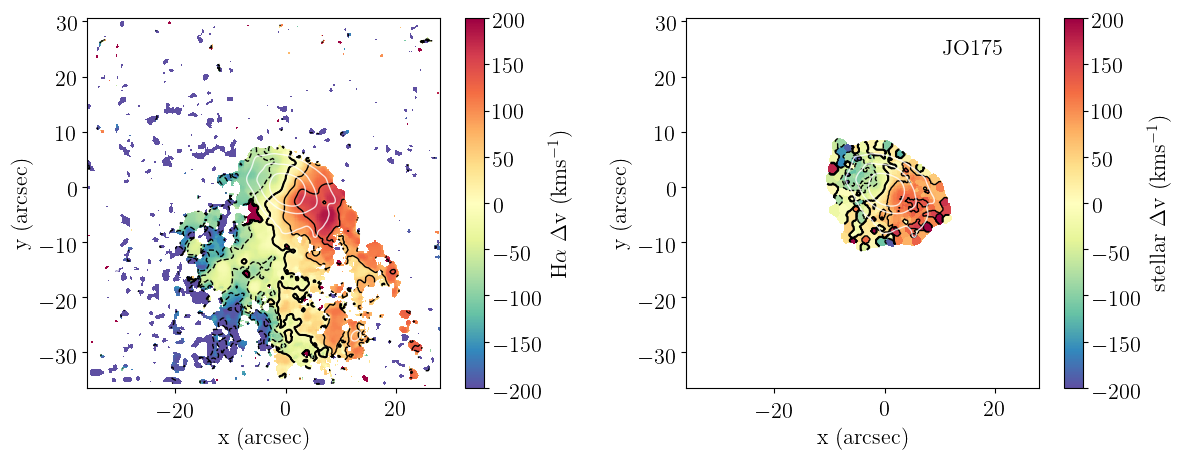}
    JO175
    \includegraphics[width=\textwidth]{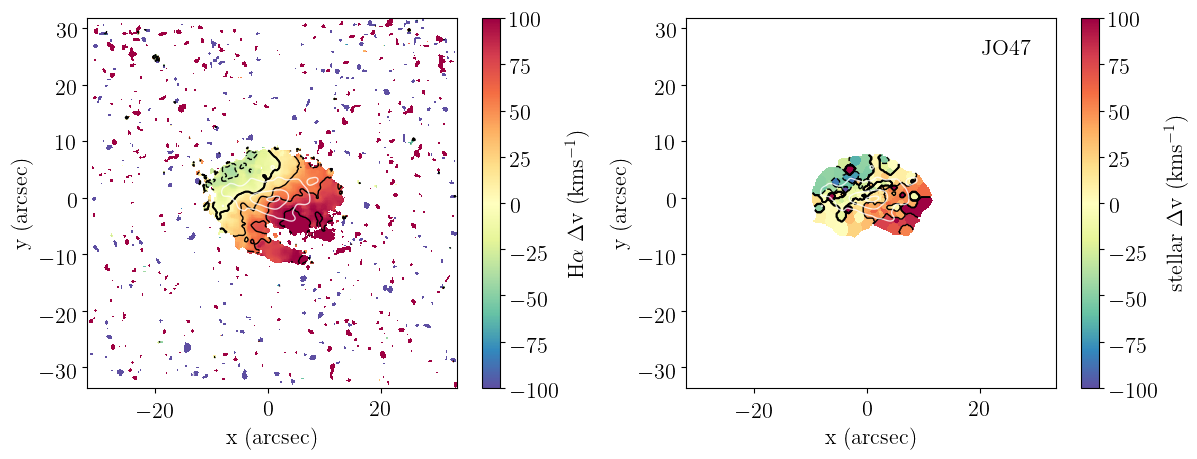}
    JO47
    \includegraphics[width=\textwidth]{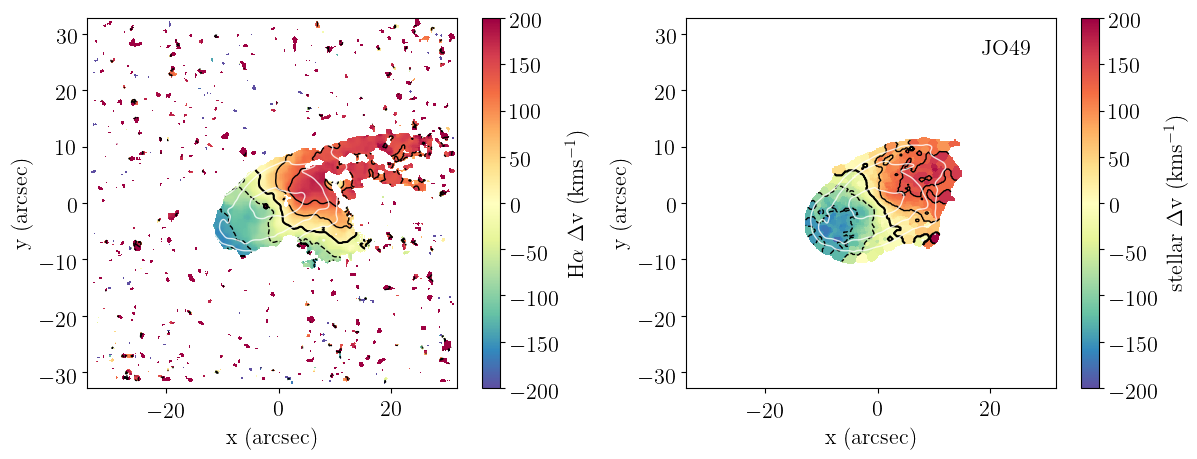}
    JO49
\end{figure*}
\begin{figure*}
    \centering
    \includegraphics[width=\textwidth]{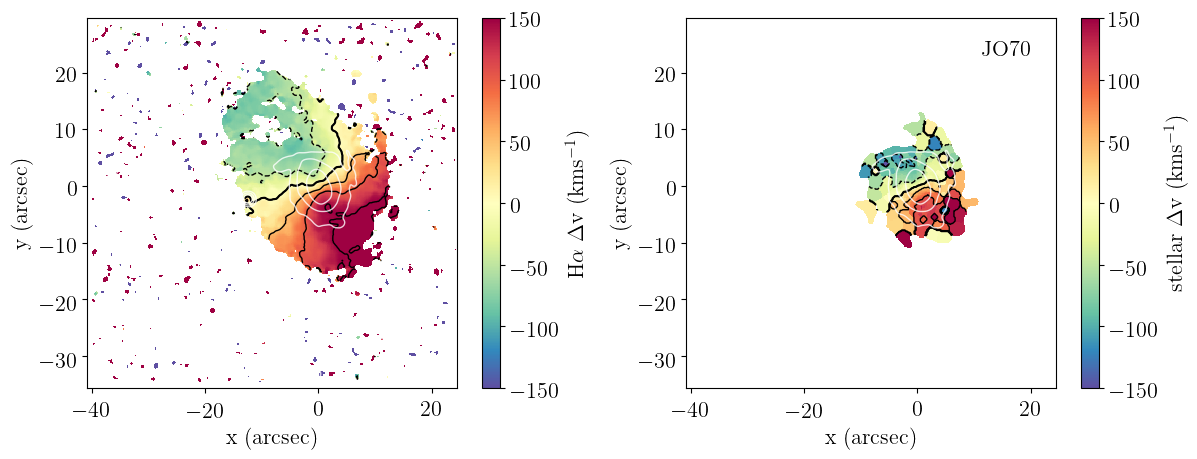}
    JO70
    \includegraphics[width=\textwidth]{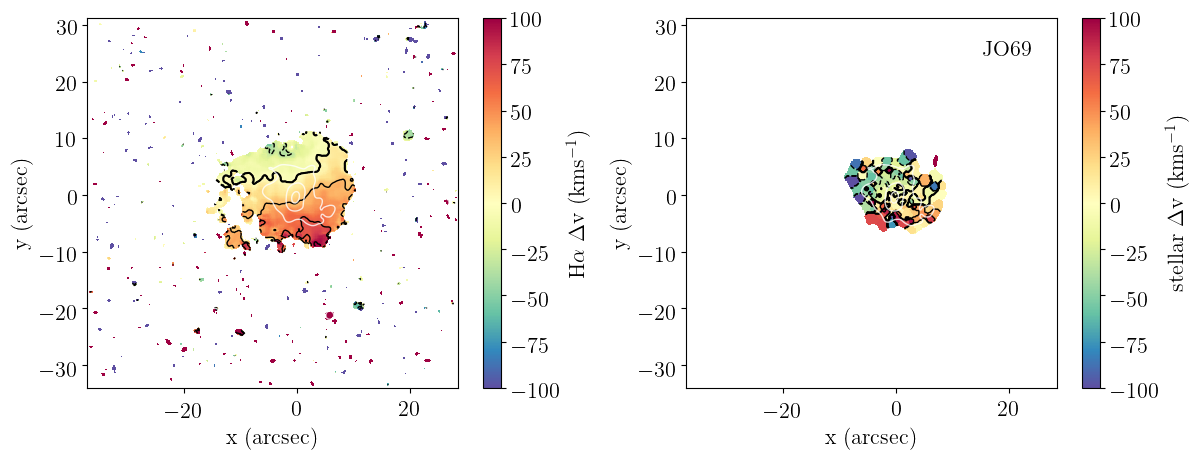}
    JO69
    \includegraphics[width=\textwidth]{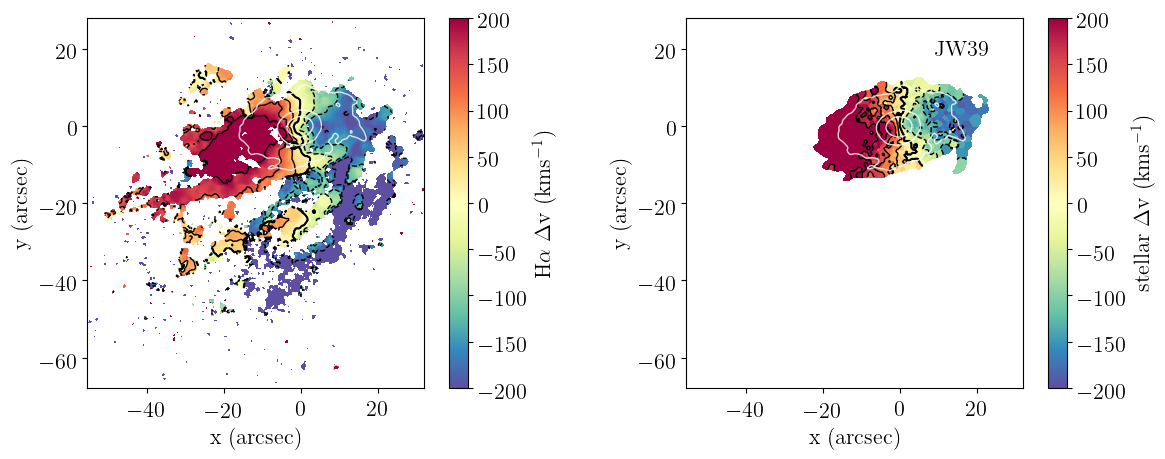}
    JW39
    \caption[]{Comparison of kinematics maps in the rest of the sample, following from Figure~\ref{fig:kin_compare_vels}. Black lines indicate iso-velocity contours on the corresponding maps. The left panels show the ionised gas velocity maps for each galaxy measured from emission line kinematics. The right panels show the stellar velocity maps measured using ppxf. White contours denote stellar continuum isophotes.}
    \label{fig:vel_compare_all}
\end{figure*}

\section{Azimuthal Diagrams for Simulated Galaxies}

\begin{figure*}
    \centering
    \includegraphics[width=\textwidth]{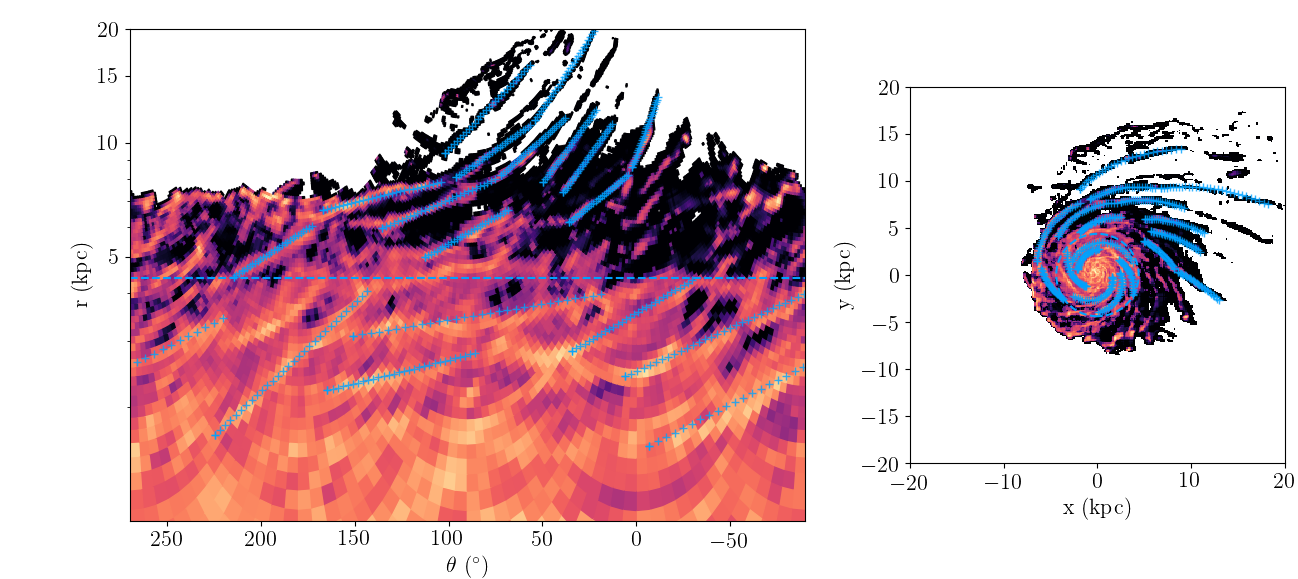}
    \includegraphics[width=\textwidth]{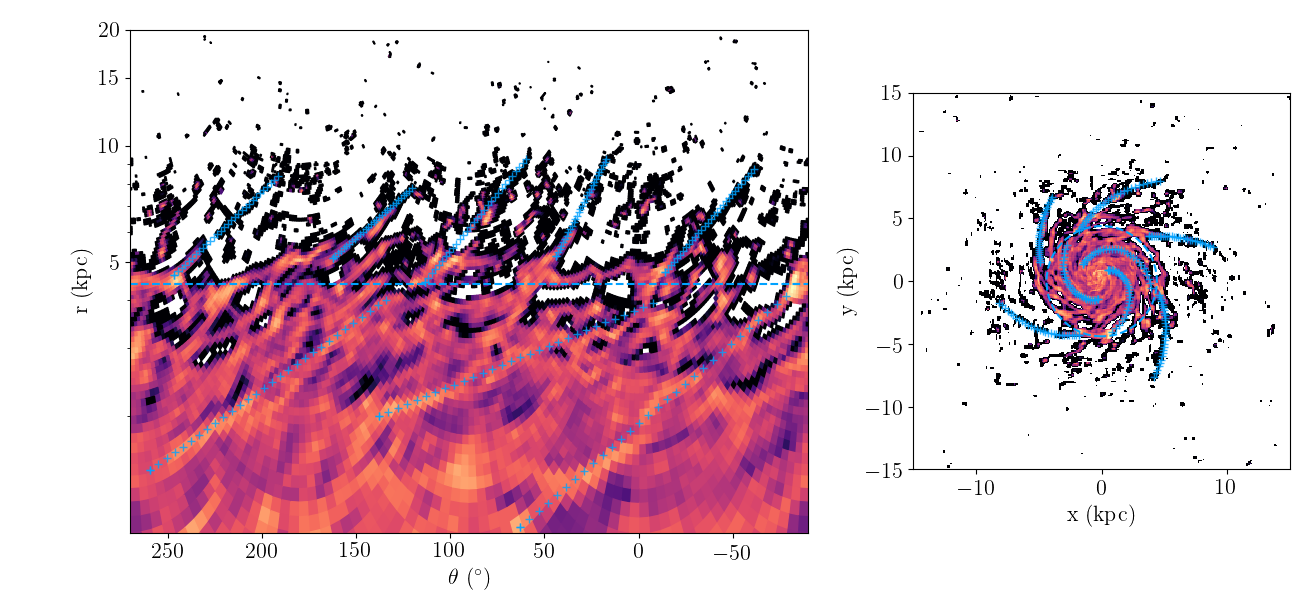}
    \caption[Plots of mean density for the "unwrapped" discs of the edge-on and face-on simulated galaxies, showing radial projected distance from the centre against azimuthal angle.]{Azimuthal plots of H$\alpha$ for the "unwrapped" discs of the edge-on stripping (\textit{Top}) and face-on stripping (\textit{Bottom}) simulated galaxies, showing radial projected distance from the centre in logarithmic scale against azimuthal angle in the left panels, alongside the original galaxy discs on the right panels. The blue dashed horizontal lines on the left panels and the dashed blue ellipses on the right mark 2 scale radii on the disc. Lines of + symbols mark spiral arm patterns identified by eye on the unwrapped figure and are shown reprojected back on the original galaxy discs. The pitch angles of these spiral arms are shown along with their average radial distance in Figure~\ref{fig:pitchangles}.}
    \label{fig:sim_unwrap}
\end{figure*}

Similarly to the azimuthal figures of the "unwrapped" discs of JO85, JO200 and P25500, we produced azimuthal figures of the "observed" simulated galaxies at the first timestep (81Myr after entering the cluster). The galaxies were rotated to be viewed face-on, before the average density was integrated in bins along the line of sight. These were then treated in the same way as the observed galaxies, reprojecting each pixel into polar space and manually tracing the spiral arms to obtain the pitch angles. The resulting plots are shown in Figure~\ref{fig:sim_unwrap}.

\section{Azimuthal Diagrams for Full Sample}\label{azimuthal_sample}

We show in this Appendix the azimuthal figures as presented in Section~\ref{sec:azimuthal} for the remainder of the sample. The values of the pitch angles measured in the inner and outer disks of each galaxy were averaged and given in Table~\ref{table:pitchangles}.

\begin{figure*}
    \centering
    JO201\\
    \includegraphics[width=0.9\textwidth]{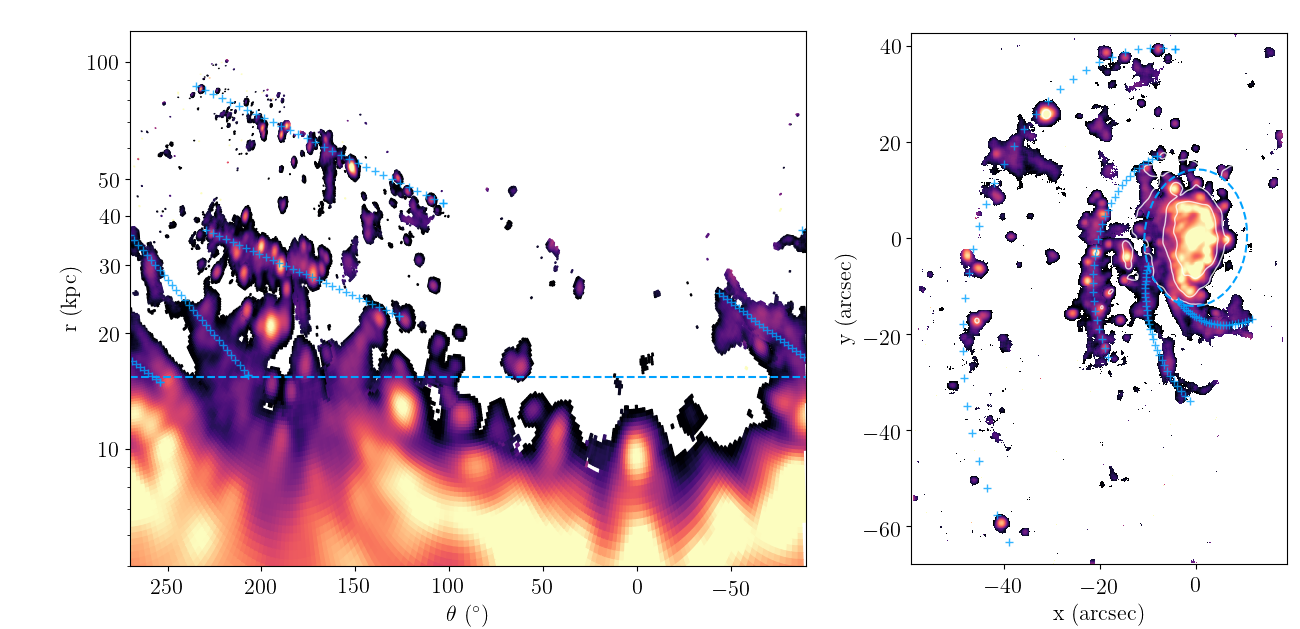}\\
    JO93\\
    \includegraphics[width=0.9\textwidth]{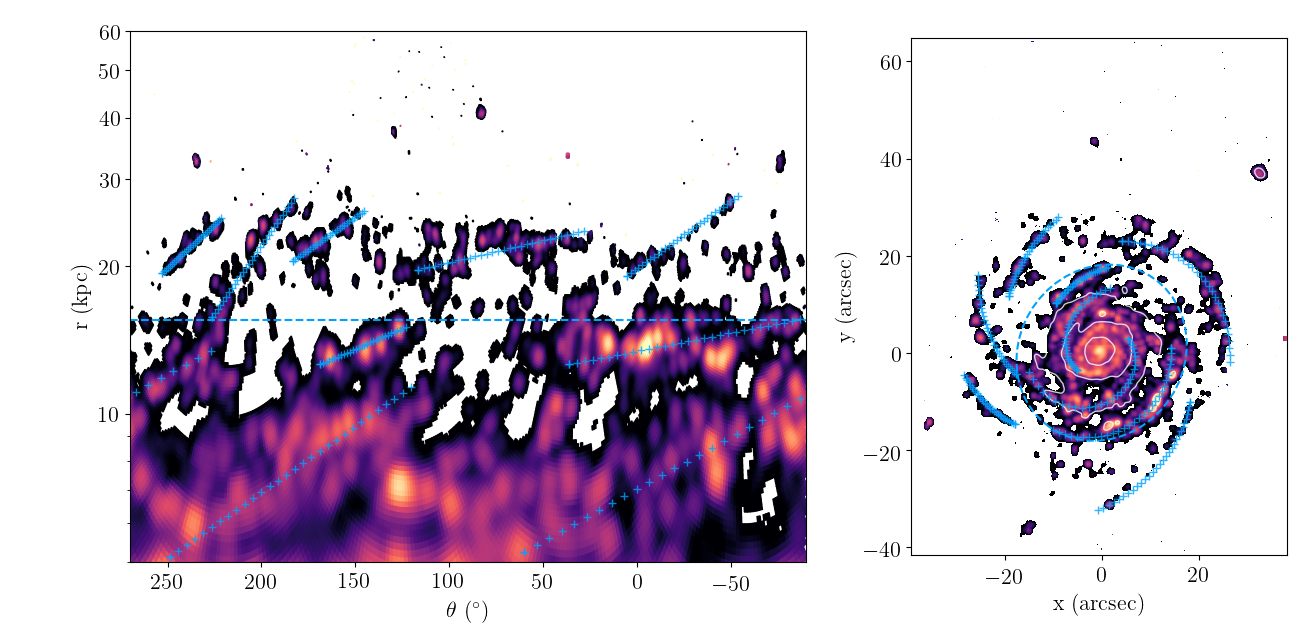}\\
    JO194\\
    \includegraphics[width=0.9\textwidth]{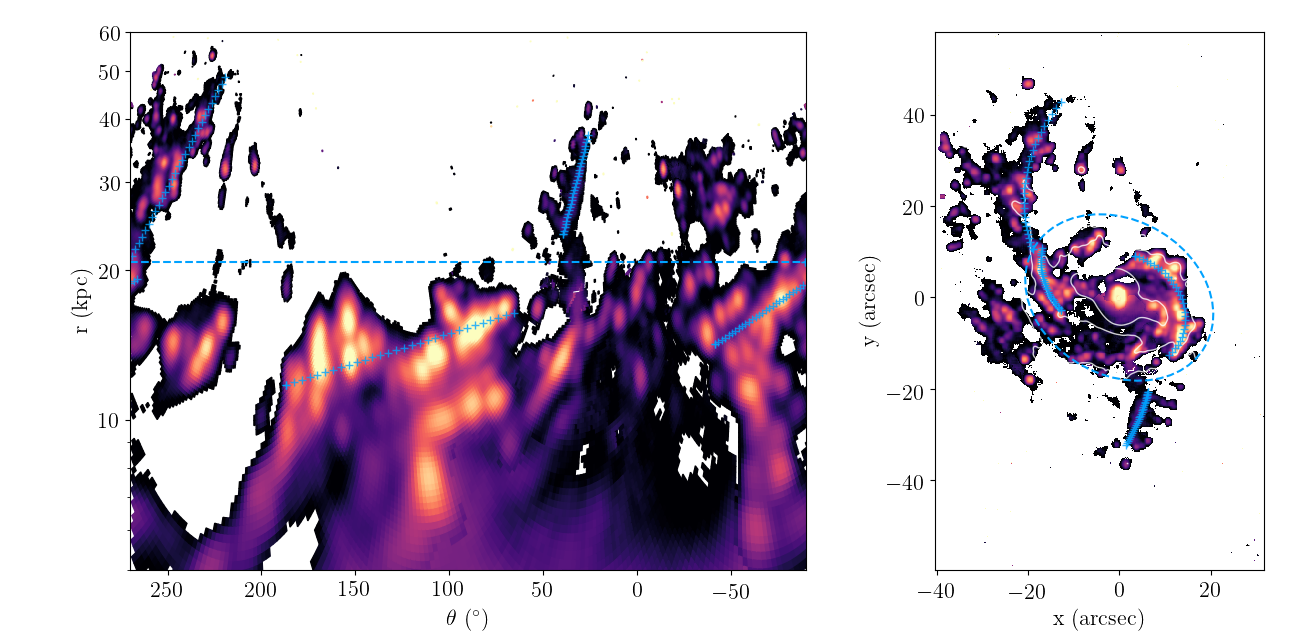}
\end{figure*}
\begin{figure*}
    \centering
    JO175\\
    \includegraphics[width=0.9\textwidth]{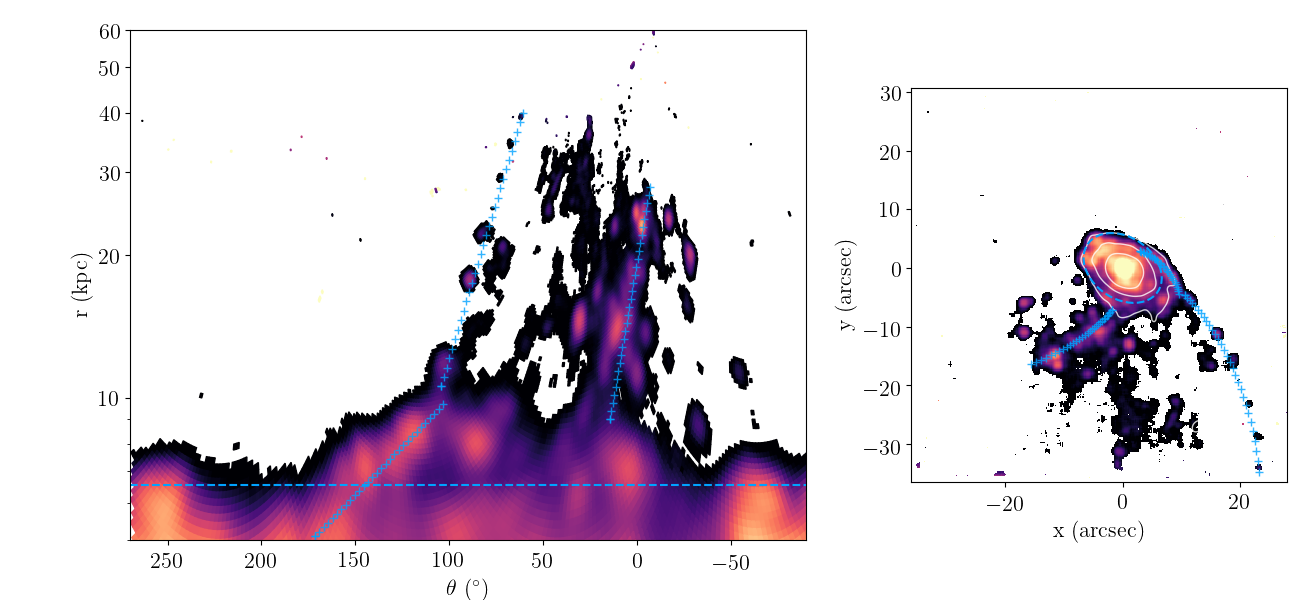}\\
    JO47\\
    \includegraphics[width=0.9\textwidth]{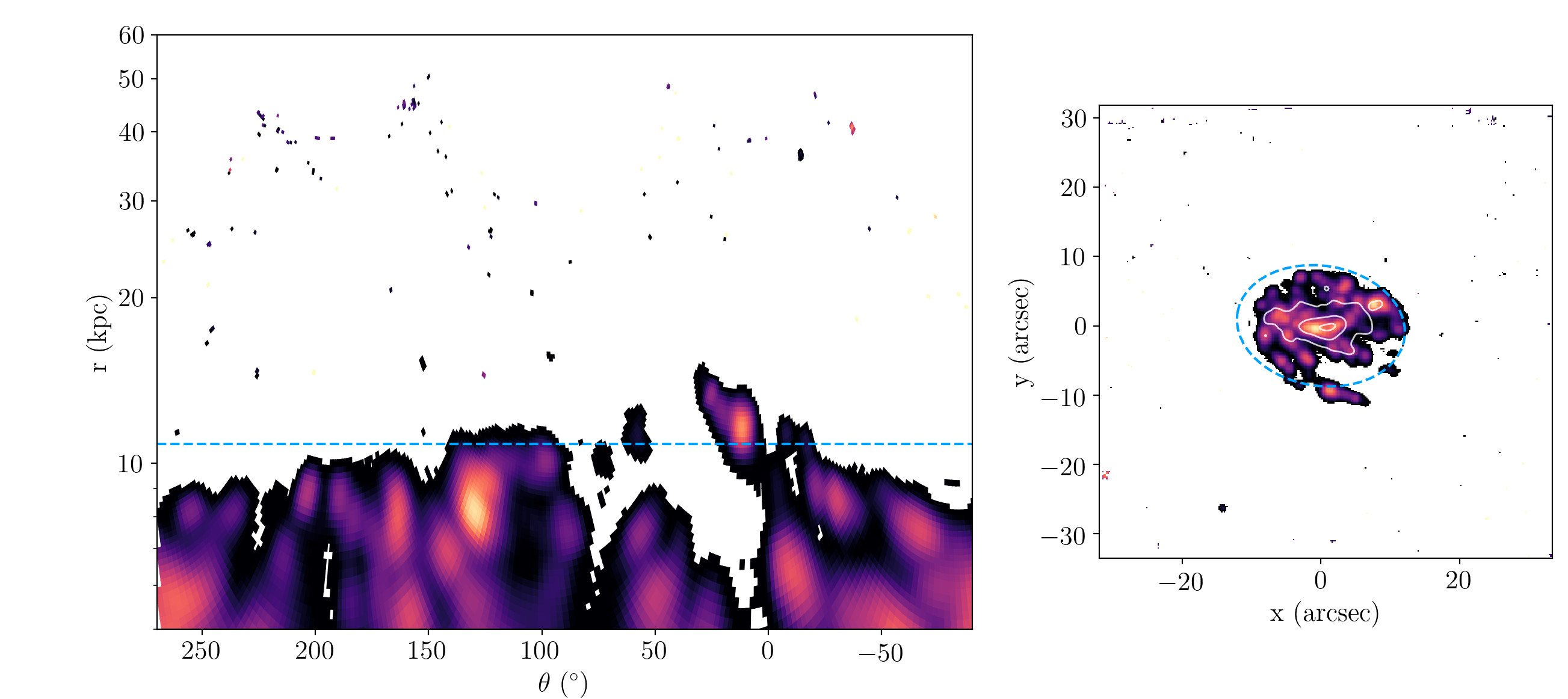}\\
    JO49\\
    \includegraphics[width=0.9\textwidth]{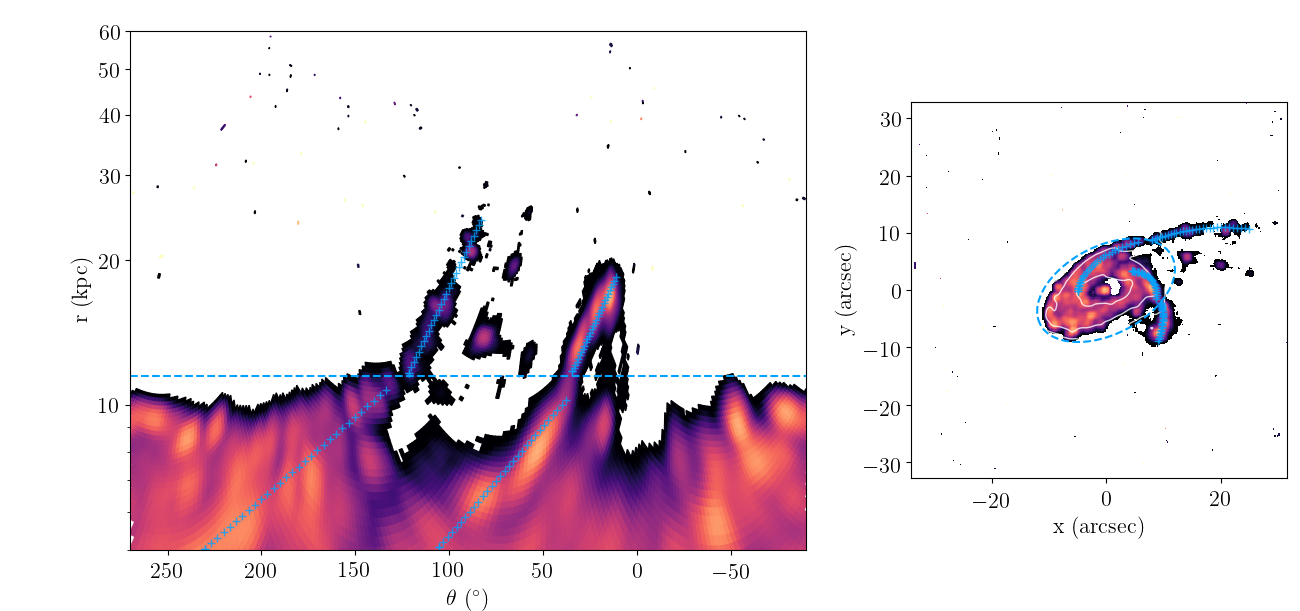}\\
\end{figure*}
\begin{figure*}
    \centering
    JO70\\
    \includegraphics[width=0.9\textwidth]{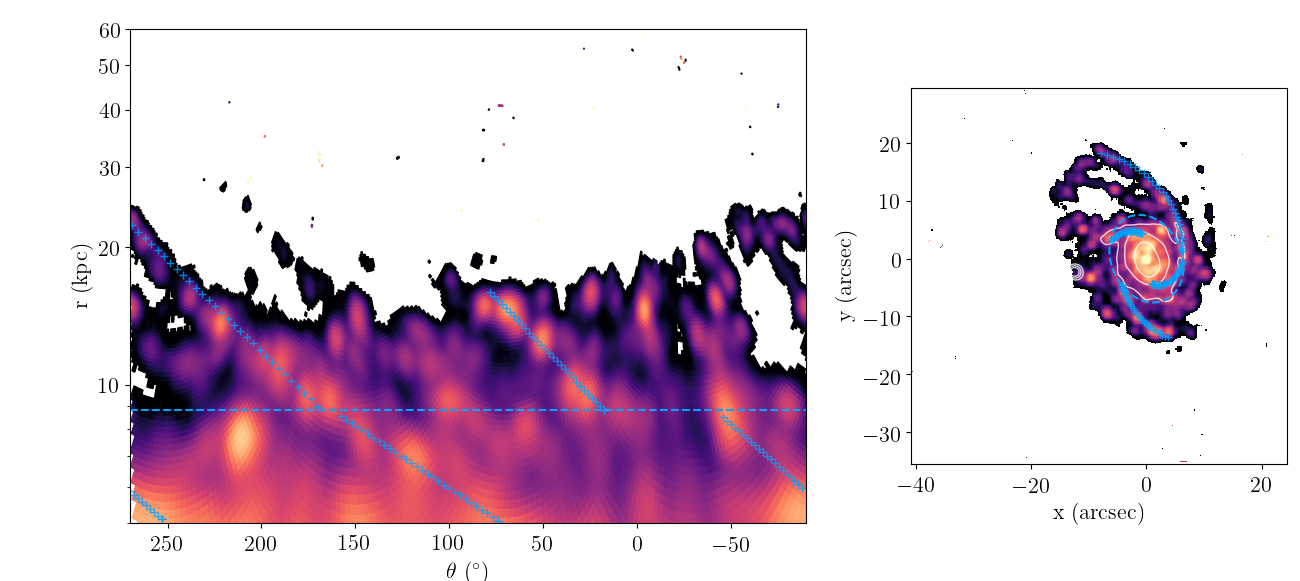}\\
    JO69\\
    \includegraphics[width=0.9\textwidth]{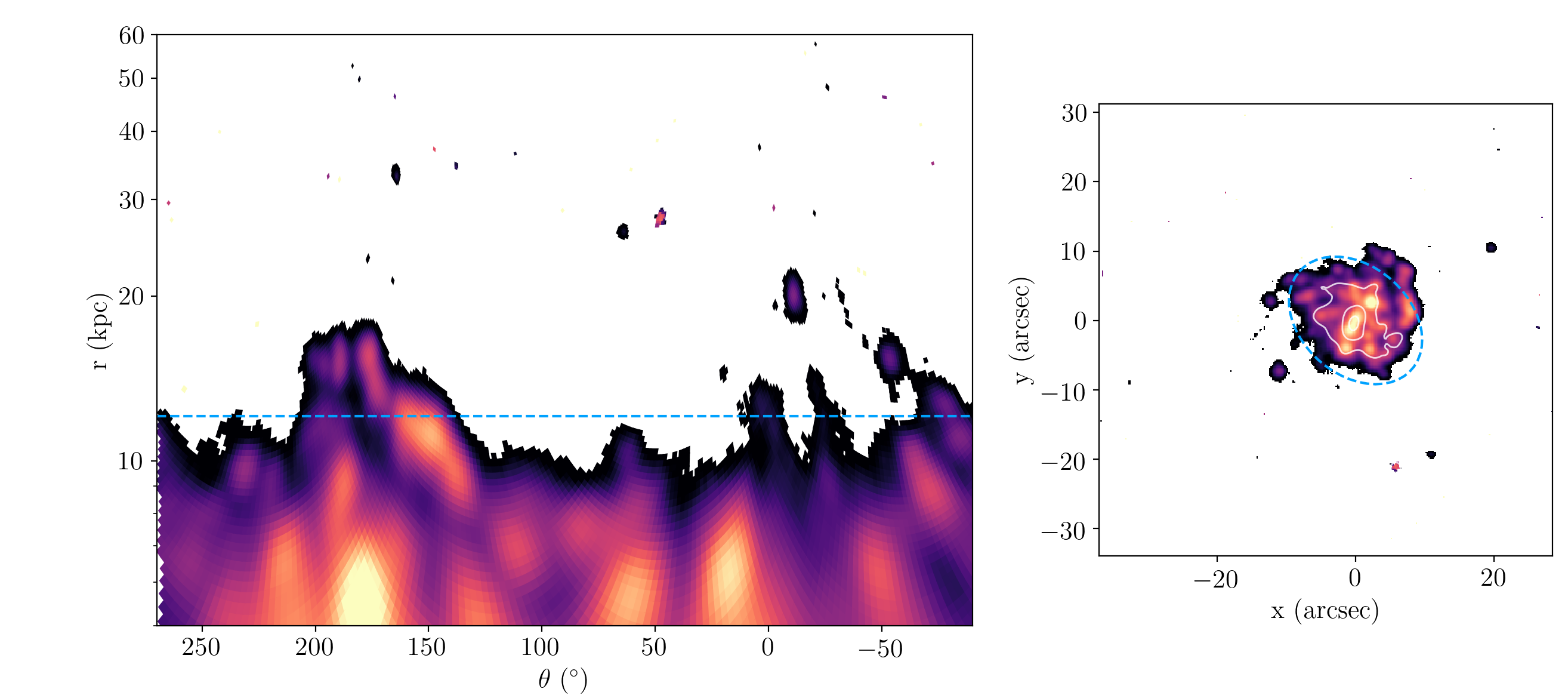}\\
    JW39\\
    \includegraphics[width=0.9\textwidth]{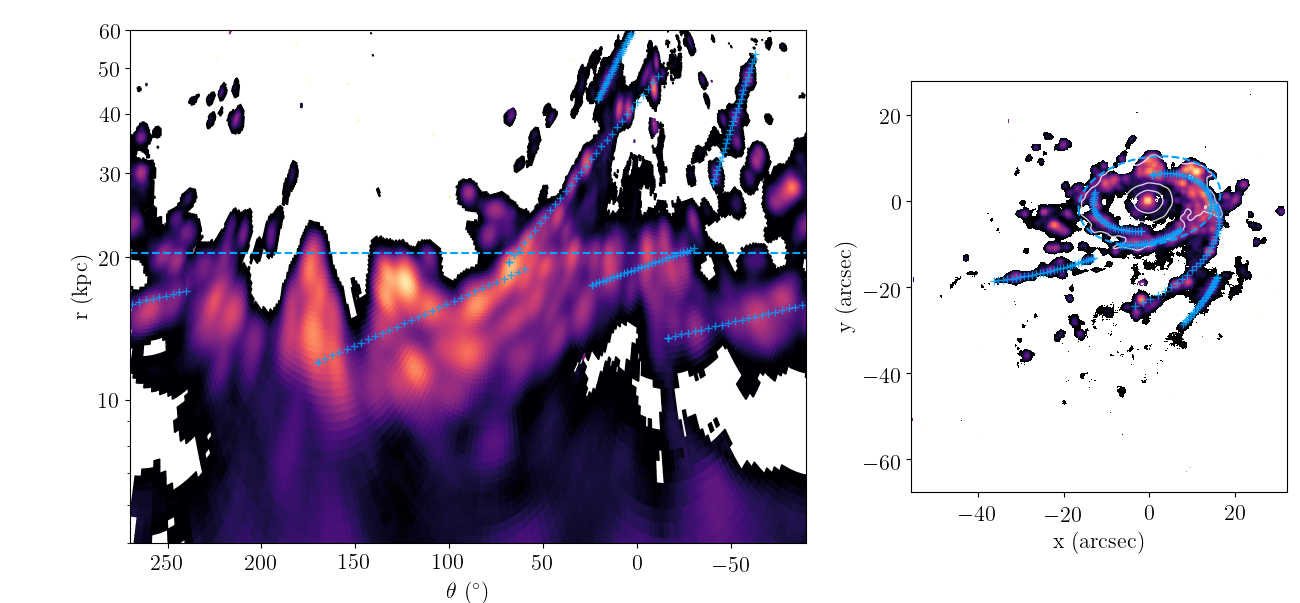}\\
\end{figure*}
\begin{figure*}
    \centering
    P21734\\
    \includegraphics[width=0.9\textwidth]{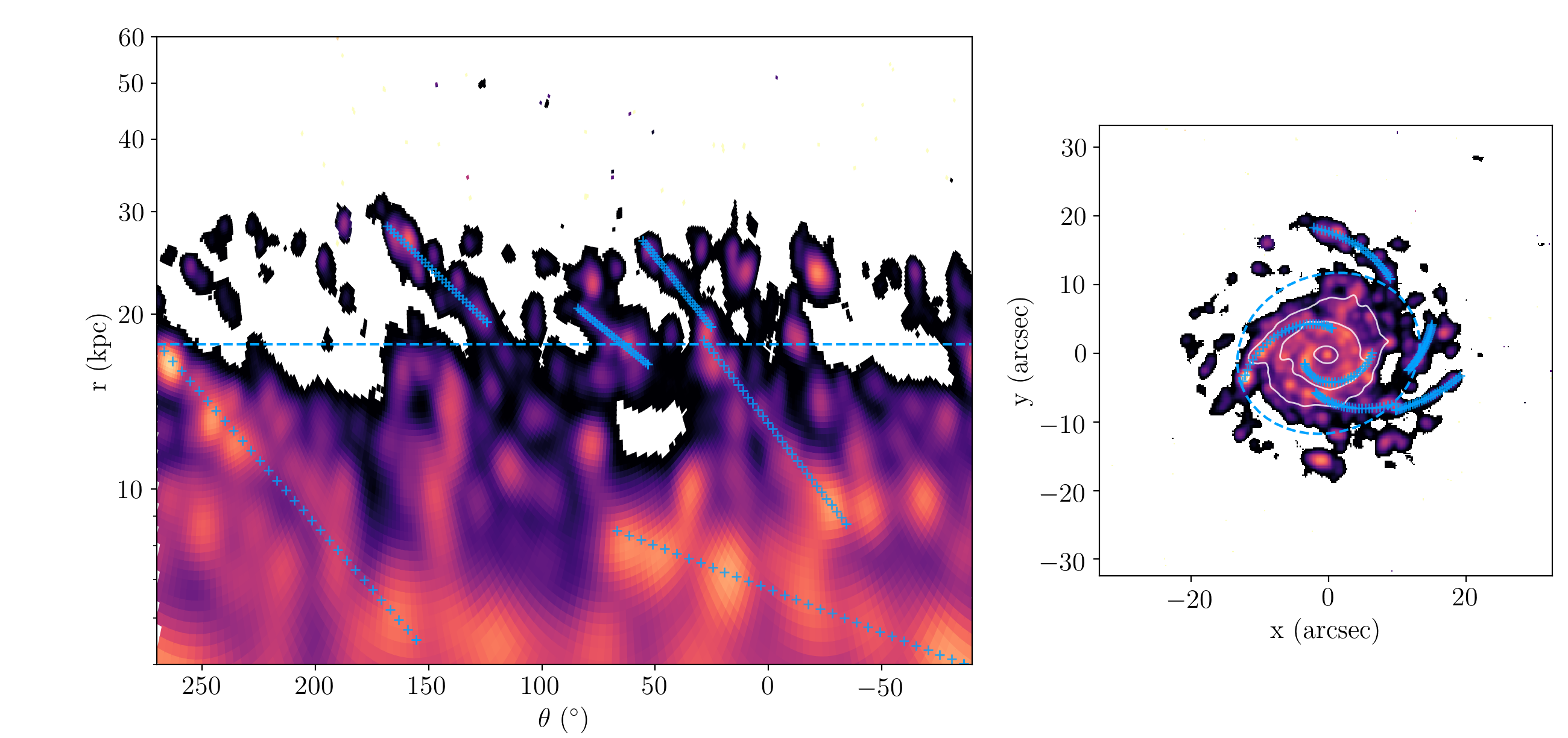}\\
    P48157\\
    \includegraphics[width=0.9\textwidth]{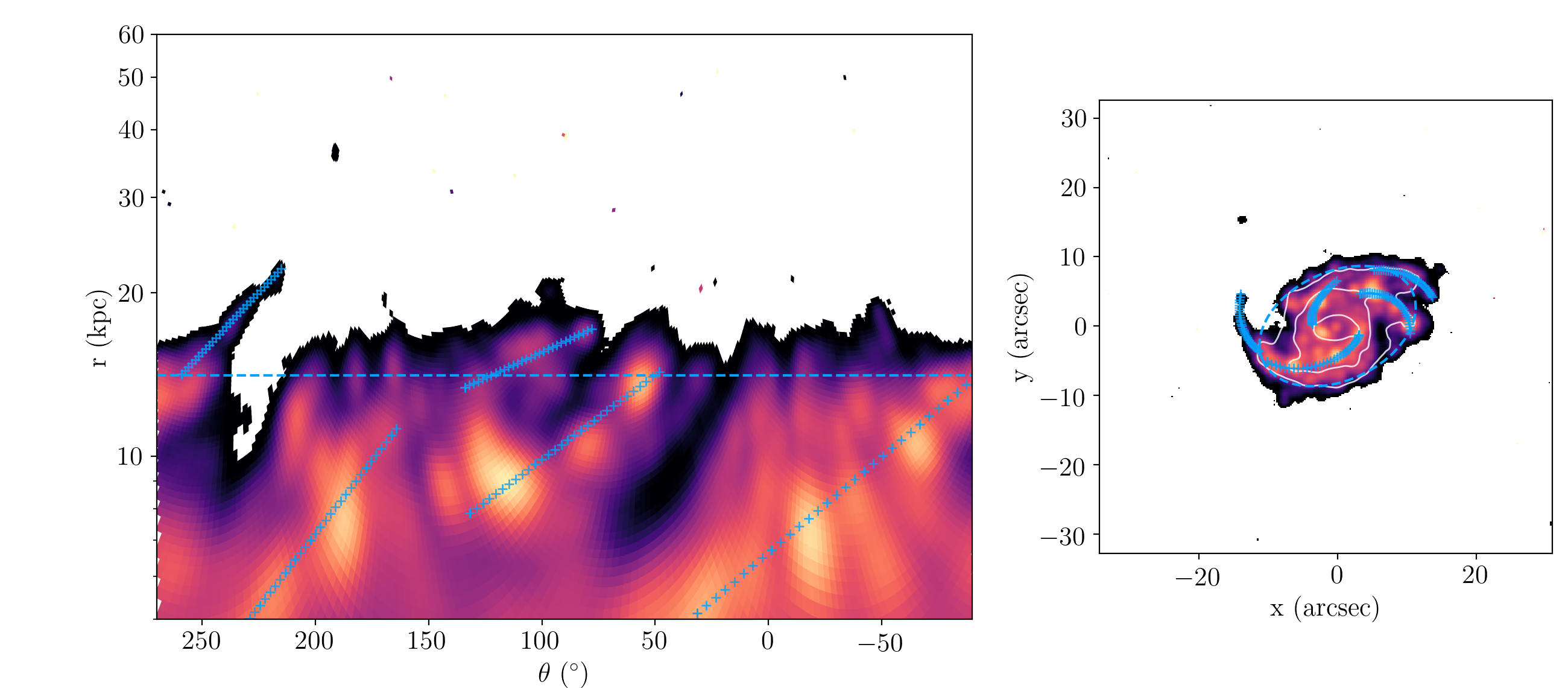}\\
    P19482\\
    \includegraphics[width=0.85\textwidth]{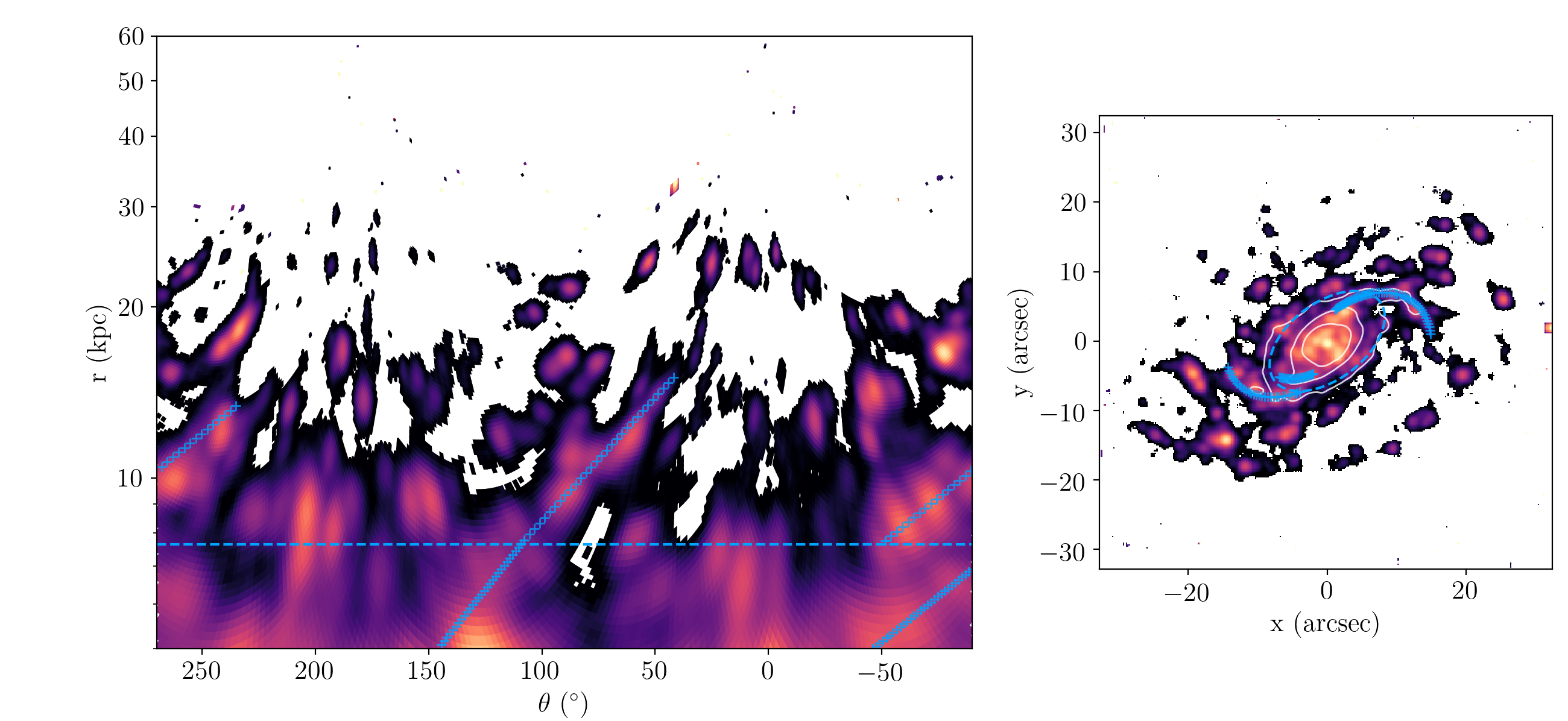}\\
\end{figure*}
\begin{figure*}
    \centering
    P63661\\
    \includegraphics[width=0.85\textwidth]{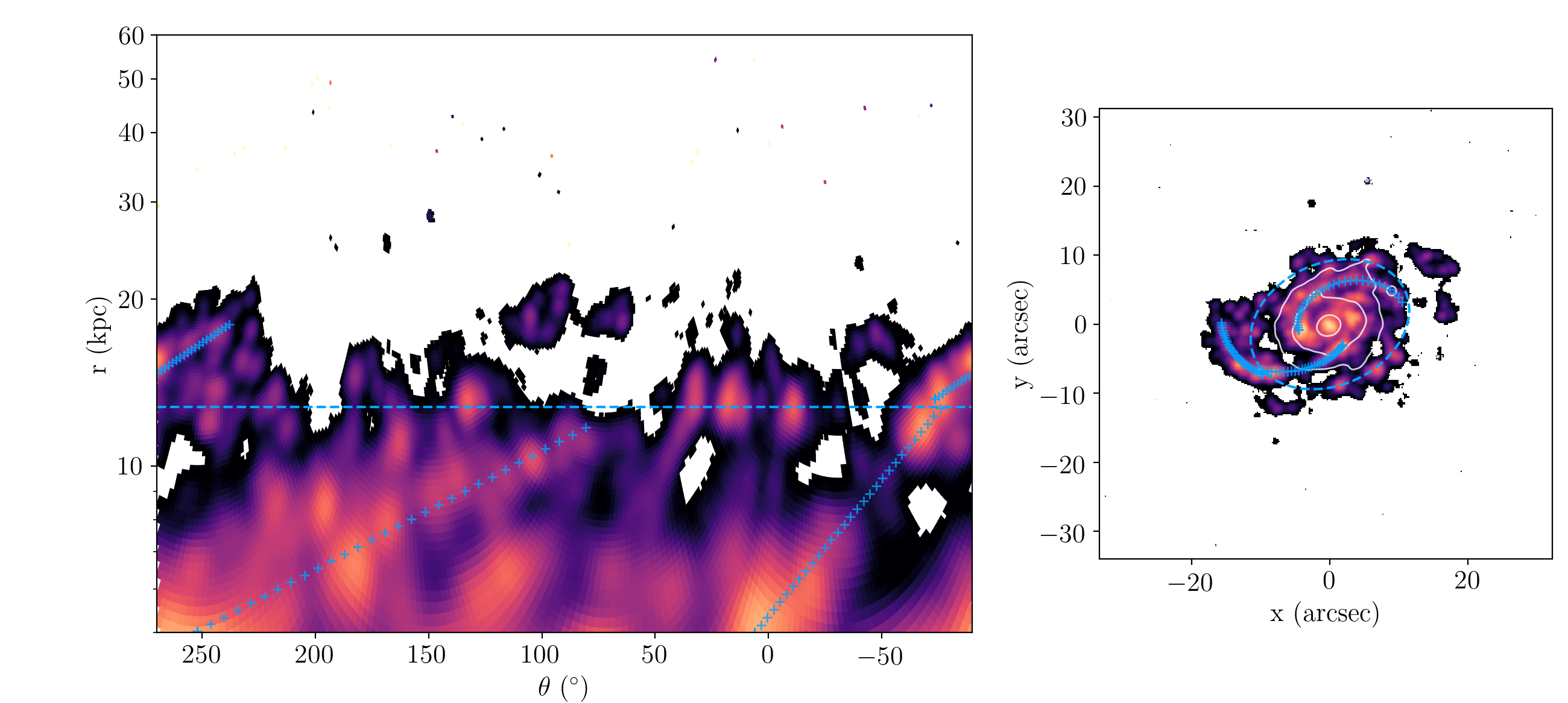}\\
    P95080\\
    \includegraphics[width=0.85\textwidth]{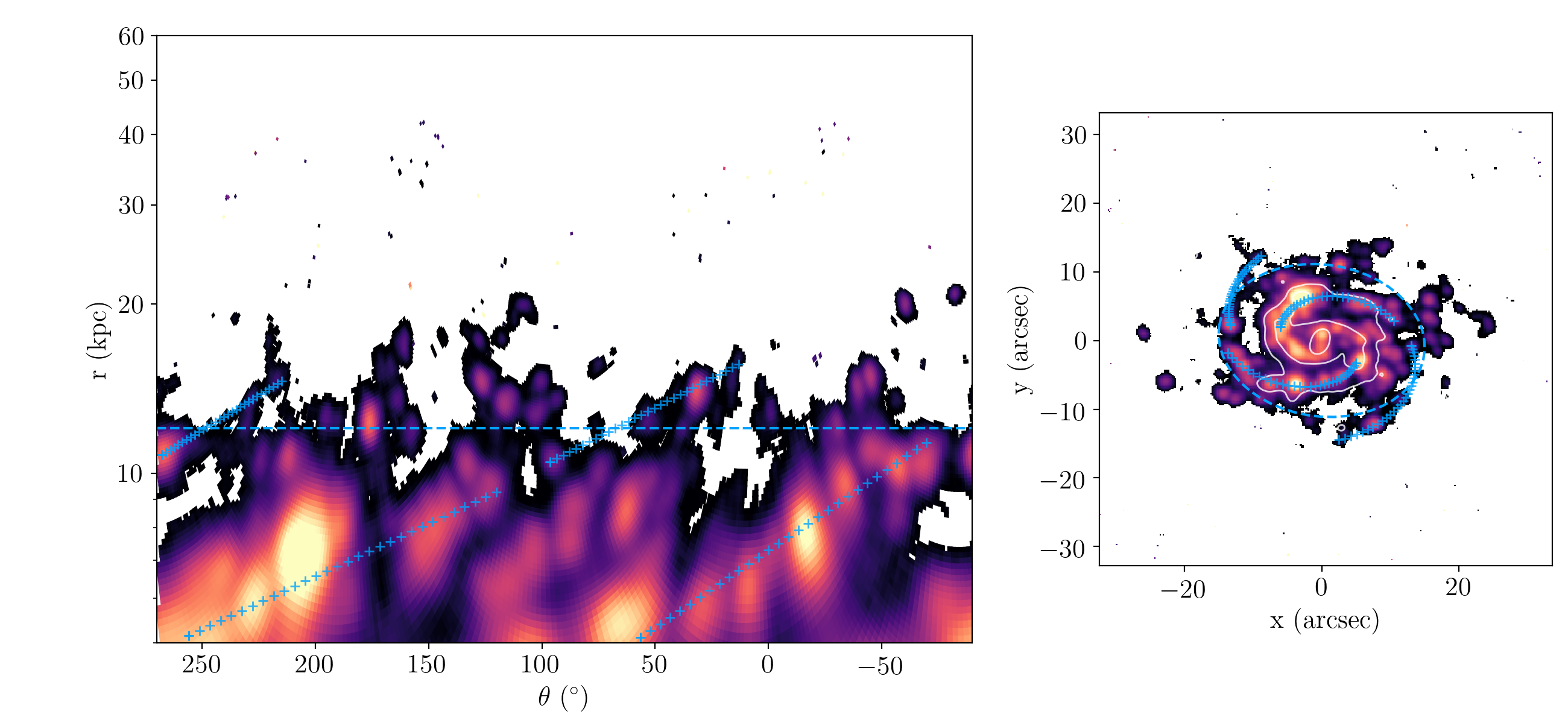}
    \caption[]{Azimuthal plots of H$\alpha$ for the "unwrapped" discs of the remainder of the unwinding sample and control galaxies, showing radial projected distance from the centre in logarithmic scale against azimuthal angle in the left panels, alongside the original galaxy discs on the right panels. The white contours on the right panels denote stellar continuum isophotes. The blue dashed horizontal lines on the left panels and the dashed blue ellipses on the right mark 2 effective radii on the disc. Lines of + symbols mark spiral arm patterns identified by eye on the unwrapped figure and are shown reprojected back on the original galaxy discs. The pitch angles of these spiral arms are shown along with their average radial distance in Figure~\ref{fig:pitchangles}.}
    \label{fig:azimuthal_compare_all}
\end{figure*}

\bsp	
\label{lastpage}
\end{document}